\documentclass[pra,aps,showpacs,twocolumn,floatfix]{revtex4-1}
\usepackage[utf8]{inputenc}
\usepackage{graphicx}
\usepackage{amsmath}
\usepackage{amsfonts}
\usepackage{braket}
\usepackage{hyperref}
\usepackage{txfonts}

\newcommand{\I}{\mathrm{i}}

\let\originalleft\left
\let\originalright\right
\renewcommand{\left}{\mathopen{}\mathclose\bgroup\originalleft}
\renewcommand{\right}{\aftergroup\egroup\originalright}
\renewcommand{\right}{\aftergroup\egroup\originalright}
\begin{document}
\title{Quantum backflow effect and nonclassicality}
\author{Francesco Albarelli}
\email{francesco.albarelli@unimi.it}
\affiliation{Quantum Technology Lab, Dipartimento di Fisica, Universit\`a degli Studi di Milano, I-20133
Milano, Italy}
\author{Tommaso Guaita}
\affiliation{Quantum Technology Lab, Dipartimento di Fisica, Universit\`a degli Studi di Milano, I-20133
Milano, Italy}
\author{Matteo G. A. Paris}
\email{matteo.paris@fisica.unimi.it}
\affiliation{Quantum Technology Lab, Dipartimento di Fisica, Universit\`a degli Studi di Milano, I-20133
Milano, Italy}
\affiliation{INFN, Sezione di Milano, I-20133 Milano, Italy}
\date{\today}
\begin{abstract}
The {\em quantum backflow effect} is a counterintuitive behavior of the
probability current of a free particle, which may be negative even 
for states with vanishing negative momentum component. Here we 
address the notion of nonclassicality arising from the backflow effect,
i.e. from the negativity of the probability current, and analyze its
relationships with the notion of nonclassicality based on the negativity
of the Wigner function. Our results suggest that backflow is linked to a
different, and in fact more restrictive, notion of nonclassicality, the
negativity of the Wigner function being only a necessary prerequisite
for its occurrence.  This hierarchical structure may be confirmed by
looking at the addition of thermal noise, which more easily destroys the
negativity of the probability current than the negativity of the Wigner
function itself.
\end{abstract} 
\maketitle
\section{Introduction}
\label{sec:intro}
The so-called {\em quantum backflow effect} consists of a
counterintuitive behavior of the probability current of a quantum free
particle in one dimension: the current may assume negative values even
for wave-packets without negative momentum components. This means that
for states with a wave-function with only positive momenta, the
probability of remaining in a certain region, e.g. $x<x_0$, may increase
with time.  This effect was earlier discovered in connection with the
discussion of the arrival-time problem in quantum mechanics
\cite{Allcock1969}, but it was studied in details only 
few years later
\cite{Bracken1994}.  In particular, a bound for the maximal fraction of
the probability that can {\em flow backwards} during a finite time
interval was found. This bound, given by the adimensional constant
$c_{bm} \approx 0.04$, is independent from the mass of the
particle and from the duration of the effect itself. Remarkably, 
this value is also independent on the Planck constant $\hbar$, 
thus suggesting backflow as an independent quantum effect.
\par
More recently, the backflow effect has attracted some more 
attention \cite{mvb10,stra12}:
improvements in the numerical estimation of $c_{bm}$ have been addressed
\cite{Eveson2005,Penz2006a} and additional bounds, analytic examples, and
connections with realistic measurements have been provided
\cite{Muga1999,Yearsley2012,Halliwell2013}. Finally, an explicit scheme
to detect backflow in a Bose-Einstein condensate has been proposed
\cite{Palmero2013}.  The effect was found
also for a particle in a linear potential \cite{Melloy1998a} and 
for a Dirac particle \cite{Melloy1998}. An analogue effect for angular
momentum has been studied as well \cite{Strange2012}.
\par
The backflow effect is an intrinsically quantum phenomenon, for which 
there is no classical analogue \cite{ole13}.  
It is intimately connected with the
fact that quantum-mechanical distributions in the phase space, e.g.
the Wigner function \cite{Wigner1932,Hillery1984}, 
are not always positive functions and thus to the idea of 
\emph{negative probability} in quantum mechanics
\cite{Feynman1987,Scully1994}.  On the other hand, 
the backflow effect appears in connection  with propagation
of a quantum particle and thus it cannot be entirely traced back to 
noncommutativity of the quantum phase space, i.e. to static
nonclassicality revealed by negativity of the Wigner function.
At the same time, despite being a dynamical effect, the 
occurrence of backflow is entirely determined by the properties 
of the initial 
quantum  state since, as we will see, it occurs for systems where the 
dynamics in the phase space is essentially classical.
\par
In the quantum statistical description of physical systems, 
the fact that quasiprobability distributions in the phase space 
may assume negative values is strongly linked 
to the notion of nonclassical states, as well as to the 
quantification of such nonclassicality
\cite{Cahill1969a,Lee1991,Lee1992,Lee1995,dod02,Vogel02,kim02,wang02,Vogel05,nk08,Vogel10,
Oli11,nk12,Fer12}. In turn, the main goal of this paper is to 
investigate whether and how the backflow of probability, i.e. negativity
of the probability current, is connected, either quantitatively or
qualitatively, to the notion of nonclassicality stemming from negativity
of the Wigner function, i.e. from phase space analysis.  Our results
indicate that quantum backflow pinpoints a different, more restrictive,
notion of nonclassicality, with the negativity of the Wigner function
being only a necessary prerequisite for the occurrence of backflow.
This picture is confirmed by looking at the effect of noise, which more
easily destroys the negativity of the 
probability current than the negativity of the
Wigner function itself.
\par
The paper is structured as follows. In Section
\ref{sec:phase_space_backflow} we review the phase space description 
of dynamics in quantum mechanics and introduce the backflow effect 
from this point of view. We also review the volume of the negative 
part of the Wigner function as a quantifier of nonclassicality, and 
briefly explore the general relationship between the two concepts.  
In Section \ref{sec:gaussians}, we explicitly explore the connection 
between negativity of probability current and negativity of the 
Wigner function for states expressed as a superposition of two
Gaussian wave packets.  In Section \ref{sec:gaussian_smoothing}, we
analyze the behavior of the backflow in the presence of noise, i.e.
under the operation of Gaussian smoothing of the Wigner function, 
and prove explicitly that negativity of the Wigner function is more
robust against noise than negativity of the probability current.
Section \ref{sec:out} closes the paper
with some concluding remarks.
\section{Phase Space Dynamics and Quantum Backflow Effect}
\label{sec:phase_space_backflow}
\subsection{Phase Space Dynamics}
\label{subsec:ps dynamics}
A pure quantum state of a particle moving 
along a line (coordinate denoted by $x$) may be described 
by its wave function in the position
representation $\psi_{t}\left(x\right)=\langle x|\psi_{t}\rangle$.
A fully equivalent representation may be also given in terms
of a phase space distribution function. In fact, the so-called Wigner
function \cite{Wigner1932} contains the full information about the 
state of the system. For a pure state the Wigner function 
is given by
\begin{align}
\label{eq:wigner_pos}
W(x,p;t) &= \frac{1}{2\pi} \int \! \mathrm{d} y \, \psi_t ^* \left(x +
\frac{1}{2}y \right)\, \psi_t \left(x - \frac{1}{2}y\right)\, 
e^{ \I p y}\,,
\\
&= \frac{1}{2\pi} \int \! \mathrm{d} q \, \phi_t ^* \left(p +
\frac{1}{2}q \right)\, \phi_t \left(p - \frac{1}{2}q\right)\, e^{- \I x
q}\,,
\label{eq:wigner_mom}
\end{align}
where the first line is the expression in terms of the position wave
function $\psi_{t}\left(x\right)$ and the second one is the equivalent
momentum representation, $\phi_{t}\left(p\right) = (2\pi)^{-1/2}\int
\mathrm{d}x\, e^{-ipx}\, \psi_t(x) \equiv\langle p|\psi_{t}\rangle$
being the momentum representation of the wave function. 
The Wigner function is a real valued function, bounded by $ |W(x,p;t)|\leq \frac{2}{\pi} $ 
and normalized. On the other hand, it may take negative values and thus 
it cannot be interpreted as a probability distribution 
in the phase space.
\par
For systems subject to a potential depending only on the coordinates,
i.e. governed by the Hamiltonian
\begin{equation}
\label{eq:hamiltonian}
H=\frac{p^2}{2m}+U\left(x\right),
\end{equation}
the Wigner function obeys the continuity equation
\begin{equation}
\label{eq:continuity}	
\frac{\partial}{\partial t}W\left(x,p;t\right) + \mathrm{div}\,
\mathbf{J}=0,
\end{equation}
where
\begin{equation}
\label{eq:wigner flow}
\mathbf{J}=\left( \begin{array}{c} J_{x}\\ J_{p}\\ \end{array}\right) 
\end{equation} 
is the Wigner function flow of the system in the phase space
\cite{Bauke2011,ole13,ole14,Skodje1989a}. This Wigner flow can be
decomposed as the product $\mathbf{J}=W\mathbf{v}$, where
$\mathbf{v}=\mathbf{J}/W$ may be interpreted as the velocity of the phase
space flow. Remarkably, for potentials at most quadratic in $x$, the 
velocity field $\mathbf{v}$ coincides with its classical analogue
\begin{equation}
\label{eq:classical ps velocity}
 \mathbf{v}=\left(\begin{array}{c} \dot{x}\\ \dot{p}\\ \end{array}
 \right)=\left(\begin{array}{c} \partial_{p} H\\ -\partial_{x} H\\
 \end{array} \right)\,.  
 \end{equation}
For this class of potentials the flow is thus Liouvillian, 
\textit{i.e.} $\mathrm{div}\,\mathbf{v}=0$, and the Wigner function 
flows in the phase space as an incompressible fluid.
\par 
Some typical quantum effects arise as a consequence of the fact that
the Wigner function can take negative values. E.g. it can be easily 
seen that in the regions where $W$ is negative the Wigner flow
$\mathbf{J}=W\mathbf{v}$ takes place in the direction opposite to
the velocity $\mathbf{v}$, which, as we have seen, gives the direction
of the classical phase space flow.
\subsection{The quantum backflow effect}
\label{sec: bf definition}
The properties illustrated in the previous Section may give rise 
to somewhat surprising results, such as the so-called quantum 
probability backflow effect.  Let us consider a one dimensional 
free particle, whose initial state is a wave packet
containing only components of positive momentum. Its wave function at
time $t=0$ is given by
\begin{equation}
\label{eq: positive momentum wave function}
\psi\left(x,0\right) =\frac{1}{\sqrt{2\pi}\hbar}
\int_{-\infty}^{+\infty}\!\!\!\! \mathrm{d}p\, e^{\frac{ipx}{\hbar}}\,
\phi\left(p\right) 
\end{equation}
where $\phi\left(p\right)$ is a function which vanishes for negative
values of $p$. In this situation the Wigner function of the particle is
entirely localized in the positive momentum half plane of the phase
space.
\par
As we have seen the Wigner flow for a free particle coincides with the
classical phase space flow, that is the one given by the velocity
\begin{equation}
\label{eq:free particle velocity}
 \mathbf{v}=\left(\begin{array}{c} \frac{p}{m}\\ 0\\ \end{array} \right).
 \end{equation}
In the positive momentum region, where our particle is localized, the
velocity is therefore always in the positive $x$ direction. However, in
points where the Wigner function takes negative values, the Wigner flow
points in the negative $x$ direction.
Let us, in particular, consider the total Wigner volume found in the
$x\geq0$ half plane in phase space. 
This is given by \begin{equation}
\label{eq:left half plane wigner volume}
\int_{-\infty}^{+\infty}\!\!\!\!\mathrm{d}p 
\int_{0}^{+\infty}\!\!\!\! \mathrm{d}x\, W\left(x,p;t\right),
\end{equation}
and it coincides with the probability of finding the particle in 
the positive  position semi-axis at a given time, that is
\begin{equation}
\label{eq:P}
P\left(t\right)=\int_{0}^{+\infty}\!\!\!\!\mathrm{d}x\, 
{\left|\psi_t\left(x\right)\right|}^{2}.
\end{equation}
By the continuity equation (\ref{eq:continuity}), the time derivative of this volume is given by the Wigner flow through the $x=0$ line in phase space:
\begin{equation}
\label{eq: wigner flow through x=0}
j \left(t\right):=\frac{d}{dt}P\left(t\right)= 
\int_{-\infty}^{+\infty}\!\!\!\!\mathrm{d}p\,\frac{p}{m}\,W\left(0,p;t\right).
\end{equation}
The expression in Eq.(\ref{eq: wigner flow through x=0}) coincides with 
the quantum mechanical probability current in the origin, i.e. 
\begin{equation}
\label{eq:quantum current}
j\left(t\right)=\frac{i\hbar}{2m}\left(\psi_t
\left(0\right)\frac{\partial\psi_t^{*}}{\partial
x}\left(0\right)-\psi_t^{*}\left(0\right)\frac{\partial\psi_t}{\partial
x}\left(0\right)\right)\,. \end{equation}
According to classical intuition, one would expect the wave packet
described above to move in the positive spatial direction with a
constant average velocity and hence the probability $P\left(t\right)$ to
increase monotonically with time, as the particle moves into the positive
position semi-axis. However, this is the case only for states which
mimic classical behavior sufficiently well, i.e. states whose
Wigner function is always positive.  Conversely, if the
Wigner function takes negative values, its phase space flow can be in
the negative direction even in the positive momentum region and
therefore, if this negative flow occurs in a sufficiently large section of
the $x=0$ line, the derivative (\ref{eq: wigner flow through x=0}) can
indeed take negative values.
\par
As a consequence, for a generic quantum state, even if in 
the limit $t\rightarrow+\infty$ the probability 
$P\left(t\right)$ globally and monotonically increases, approaching 
the limiting value $P\left(t\right)=1$, there may exist time 
intervals in which it is a locally decreasing function 
of time. The particle thus appears to return towards 
the negative semi-axis. In order to quantify the backflow 
effect, one may consider the maximum amplitude of such temporary 
decrease of the probability density, i.e.
\begin{equation}
\label{bf definition}
\beta \left[\psi\right]:=\left|\inf_{t_{1}<t_{2}} 
\left[P\left(t_{2}\right)-P\left(t_{1}\right)\right]\right|\,.
\end{equation}
The increase in probability over a time interval $\left(t_{1},
t_{2}\right)$ (the most negative values of which we must find 
to compute backflow) can be expressed in terms of the phase 
space flow in Eq. (\ref{eq: wigner flow through x=0}) as follows 
\begin{equation}
\label{eq:flux function}
F\left(t_{1},t_{2}\right):=P\left(t_{2}\right)-
P\left(t_{1}\right)=\int_{t_1}^{t_2}\!\!\!\!\mathrm{d}t
j\left(t\right)\,. 
\end{equation}
Upon considering the incompressible fluid nature of the Wigner flow, 
one may define a natural motion of phase space points so that this 
motion has velocity given by the field $\mathbf{v}$: a point
initially in $\left(x,p\right)$, after a time interval $t$ 
is mapped to
\begin{equation}
\label{eq: ps motion}
\varphi_t\left(x,p\right)=\left(\begin{array}{c}x+
\frac{p}{m}t\\p\\ \end{array}\right)\,.
\end{equation}
Because of the incompressible nature of the flow, the Wigner density
remains constant along this motion, that is 
\begin{equation}
\label{eq: time evolution of W}
W\left(x,p;t\right)=W\left(\varphi_{-t}\left(x,p\right);0\right)\,.
\end{equation}
Using this result we can express function (\ref{eq:flux function}) as
\begin{align}
\label{eq:F as integral on Omega}
F\left(t_{1},t_{2}\right)&=
\int_{\mathcal{R}}\!\!\mathrm{d}x\,\mathrm{d}p\,W\left(x,p;t_2\right)
- \int_{\mathcal{R}}\!\!\mathrm{d}x\,\mathrm{d}p\,W\left(x,p;t_1\right)
 \\ &=
\int_{\varphi_{-t_2}(\mathcal{R})}\!\!\mathrm{d}x\,
\mathrm{d}p\,W\left(x,p;0\right)
- \int_{\varphi_{-t_1}(\mathcal{R})}\!\!\mathrm{d}x\,
\mathrm{d}p\,W\left(x,p;0\right)
 \\ &=
\int_{\Omega}\!\!\mathrm{d}x\,\mathrm{d}p\,W\left(x,p;0\right)\,,
\end{align}
where $\mathcal{R}$ is the $x\geq0$ half-plane and the region 
$\Omega=\mathbf{\varphi}_{-t_{1}}\left(\mathcal{R}\right)
\setminus\mathbf{\varphi}_{-t_{2}}\left(\mathcal{R}\right)$ is 
an angular sector in the phase space. In polar coordinates $\Omega$ 
is defined by
\begin{equation}
\label{eq:angular sector}
\frac{\pi}{2} + \arctan \left( \frac{t_1}{m} 
\right) \leq \phi \leq \frac{\pi}{2} + \arctan 
\left( \frac{t_2}{m} \right)\,,
\end{equation}
and no constraint on the radial coordinate.
The increase in probability over the time interval
$\left(t_{1},t_{2}\right)$ may be thus seen as the flow of the Wigner
volume initially (at $t=0$) in the region $\Omega$ into the $x\geq0$
half-plane.  If there exists at time $t=0$ a sector $\Omega$ in which
the Wigner function has negative integral, then there is also a time
interval in which this probability increase is actually negative and the
state shows the backflow effect. See \cite{Werner1988}
for a detailed 
analysis of integrals of the Wigner function on angular 
sectors in phase spaces.
\subsection{Quantum backflow and nonclassicality}
\label{sec: bf and nonclassicality}
The backflow effect cannot be observed for a particle moving
according to the classical laws of motion. In this sense its
occurrence is a manifestation of the genuine quantum nature 
of the state under investigation. In the previous Section, we 
have seen how negativity of the Wigner function is a prerequisite
to observe negativity of the probability current, and a question 
arises about the general connection between the two notions of 
nonclassicality.
\par
A common approach to the notion of nonclassicality involves
the noncommuting nature of quantum canonical variables, which
implies the existence of an entire family of $s$-ordered 
quasidistributions in the phase space \cite{Cahill1969a}. 
Singularity, or negativity, of one or more quasidistributions
implies that the methods of classical statistics fail to 
describe the properties and the phenomenology of a given state, 
and is thus taken as a signature of nonclassicality \cite{Lee1991}. 
The most fundamental definition of nonclassicality is based on 
the Glauber-Sudarshan $P$ function, whereas negativity of the 
Wigner function, besides being measurable experimentally
\cite{ian95,kon96,lda00,lvo01,all09}, captures 
the nonclassical features involved in quantum interference \cite{h74} and 
has been recognized as 
a resource for quantum computation \cite{Mari2012,Veitch2013,Pashayan2015}.
More generally, the presence of negative values in the Wigner 
function of a quantum state have been recognized as a sufficient condition 
for nonclassicality. In particular, the volume of the negative 
part has been introduced as a nonclassicality measure \cite{Kenfack2004}.
\par
We will use the actual volume of the negative part (not its double 
or its normalized versions, sometimes used in the literature), 
which is defined as
\begin{equation}
\label{eq:Delta}
\Delta [\psi] = \frac{1}{2} \int\!\!\!\int\!\! \mathrm{d}x\, \mathrm{d}p 
\, \left[ \Big|W(x,p;t)\Big| - W(x,p;t) \right]\,.
\end{equation}
If we choose $t_1$ and $t_2$ as the time interval corresponding 
to the minimum in Eq. (\ref{bf definition}), 
then $-F\left(t_1,t_2\right)$ is equal to 
the backflow measure of the state $\beta \left[\psi \right]$. 
In this way we may identify the Wigner negativity volume $\Delta$ 
as an upper bound to the backflow: if we denote
by $V_{\Omega}^{+}$ ($V_{\Omega}^{-}$) the volume of the positive
(negative) part of the Wigner function on the sector $\Omega$ then,
recalling equation (\ref{eq:F as integral on Omega}), we may 
write the following inequality
\begin{equation}
\label{bf inequality}
\beta \left[\psi \right] = - \left( V_{\Omega}^{+} - V_{\Omega}^{-}
\right) \leq V_{\Omega}^{-} \leq \Delta\,.  \end{equation}
This confirms that nonclassicality as defined by Eq. 
(\ref{eq:Delta}) is a necessary condition for backflow. Moreover, 
a question arises on whether a more precise quantitative
relation exists between  $\beta$ and $\Delta$. In order to check 
whether this is the case, we consider an explicit example and 
analyze in some details the two quantities for superpositions 
of Gaussian wave-packets.
\section{Superpositions of Gaussian States}
\label{sec:gaussians}
\subsection{Quantum backflow for superpositions of Gaussian States}
The quantum backflow effect is not observed in states with a
sufficiently classical behavior, such as those with a Gaussian 
wave-function. However, it may easily arise by taking quantum 
superpositions of such semiclassical states, which provide 
a natural case study. In particular, we are going to consider 
the superposition of two Gaussian momentum wave-packets of 
width $\sigma$ centered on different positive momenta. An overview
of quantum backflow for such states may be found 
in \cite{Yearsley2012}. For $\sigma\rightarrow\infty$ one recovers 
a superposition of two plane waves with different momenta, which is 
the simplest example of a state presenting backflow 
\cite{Bracken1994,Yearsley2013}, though it does not correspond to 
a physical state. In the following, we analyze the backflow 
for a general normalized superposition
with complex coefficients of two Gaussian wave packets.
These state are an example of the \emph{Gaussian cat states} 
\cite{Nicacio2010}, introduced as a generalization of the so-called 
cat states often studied in quantum optics \cite{Yurke1986,dodonov2003}.
\par
Our focus is
not on a systematical analysis of the effect in the whole range of 
physical parameters. Rather, our main goal is to compare backflow and
nonclassicality in some relevant settings. To this aim, we are 
interested in finding a state which gives a local (in the parameter 
space) maximum of the backflow and to study the states in the
neighboring region of the parameter space.
\par
Upon switching to natural units (i.e. $\hbar=1$) and choosing a particle
with unit mass, $m=1$, we consider states with the following initial 
momentum representation 
\begin{equation}
\phi_{0}\left(p\right)=N\left[e^{-\left(p-p_{0}-\delta\right)^2\sigma^2}
+\alpha e^{i \theta} e^{-\left(p-p_{0}\right)^2\sigma^2}\right]\,, 
\label{eq:double gaussian state}
\end{equation}
where all the parameters are real numbers.
The normalization condition fixes the value of $N$ in terms of the other
parameters as follows 
\begin{equation}
N\left(\sigma,p_{0},\delta,\alpha,\theta\right)=
\left(\frac{2\sigma^2}{\pi}\right)^{\frac{1}{4}}
\left(1+\alpha^{2}+2e^{-\frac{1}{2}\delta^{2}
\sigma^{2}}\alpha\cos\theta\right)^{-\frac{1}{2}}.\label{eq:normalisation}
\end{equation}
The time evolved wave function, its expression in position
representation and the time dependent probability current in the origin
can be calculated analytically, but their expressions are somewhat
cumbersome and not particularly instructive: we are not reporting 
their explicit expressions here. One can see, however, that these 
quantities can be more conveniently expressed in terms of the 
following rescaled adimensional parameters:
\begin{align}
\label{eq:rescaled_parameters}
\tilde{p}_{0} =\sigma p_{0} \qquad 
\tilde{t} =\frac{t}{\sigma^{2}} \qquad 
\tilde{\delta} =\sigma\delta\,.
\end{align}
With this choice, the current $j\left(\tilde{t}\right)$ can be 
expressed as the product of a dimensional factor 
$\frac{1}{\sigma^2}$ with an adimensional oscillating
function of the remaining parameters $\tilde{j}\left(\tilde{t};
\tilde{p}_{0},\tilde{\delta},\alpha,\theta\right)$. Upon applying a 
change of variables to the integral in Eq. (\ref{eq:flux function}) 
we obtain:
\begin{equation}
\label{eq:flux with rescaled parameters}
F\left(t_{1},t_{2}\right)=\int_{\tilde{t}_{1}}^{\tilde{t}_{2}}
\!\!\!\mathrm{d}\mbox{\ensuremath{\tilde{t}}}\, j\left(\tilde{t}\right)
\end{equation} with $\tilde{t}_{k}=\sigma^{2}t_{k}$, $k=1,2$, 
from which it is apparent that the width $\sigma$ only changes 
the size of the time interval in which backflow is observed, while 
the value of the backflow itself only depends on the adimensional 
parameters $\tilde{p}_{0}$, $\tilde{\delta}$, $\alpha$ and $\theta$. 
This is in agreement with Ref. \cite{Bracken1994}, where it is emphasized that 
the duration of the backflow effect can be changed arbitrarily.
However, this extra degree of freedom may be useful if we want to
consider states at fixed energy. Indeed, if we want to maximize the
backflow at fixed energy $E$, we can minimize the flux \eqref{eq:flux
with rescaled parameters} as a function of $\tilde{p}_{0}$,
$\tilde{\delta}$, $\alpha$ and $\theta$, and then choose the appropriate
value of $\sigma$ to obtain a state with a given value of energy $E$.
\par
Of course these states do not strictly fulfill the requirement of not
containing negative momenta. On the other hand, the total volume of the wave
function localized on the negative semi-axis in momentum representation
can be arbitrarily reduced by taking a Gaussian centered on a positive
momentum sufficiently larger than its width. 
Indeed, by taking in only values of $\tilde{p}_{0}$ larger than $3$, the
negative volume is of the order of $10^{-9}$, a value corresponding to
irrelevant effect on the backflow (an explicit numerical check has been 
performed).
\par
\begin{figure}[h!]
\begin{center}
\includegraphics[width=8cm]{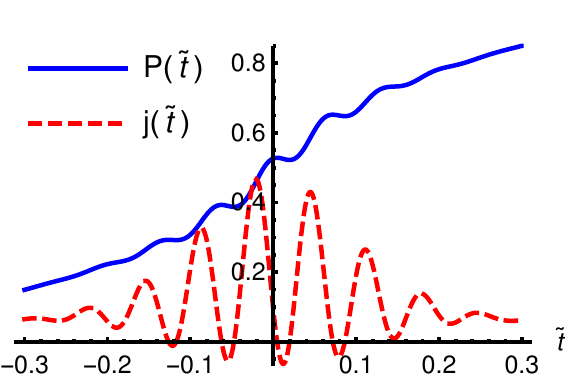}
\end{center}
\caption{(Color online) 
The probability $P\left(\tilde{t}\right)$ (solid blue curve) 
and the current $j\left(\tilde{t}\right)$ (dashed red curve) for a 
superposition of Gaussian wave-packets with $\alpha=2$, 
$\tilde{\delta}=11$, $\tilde{p}_{0}=3$, $\theta=\frac{\pi}{4}$, 
and $\sigma=10$. See Eq. (\ref{eq:double gaussian state}) for details.
The value of the current $j\left(\tilde{t}\right)$ 
is multiplied by $10$ in the figure in order to appreciate its behavior.}
\label{fig: probability}
\end{figure}

Fig. \ref{fig: probability} shows the probability 
$P\left(\tilde{t}\right)$ and current $\tilde{j}\left(\tilde{t
}\right)$ for a given superposition of two Gaussian wave-packets. 
As it is apparent from the plot, 	the time intervals where 
probability decreases coincide with the negative regions of 
the current. According to Eq. \eqref{eq:flux function} and 
since the probability $P\left(\tilde{t}\right)$ is known analytically, 
the backflow 
may be easily computed if we know the time interval which 
contains the most negative peak of the current.
However, this involves finding the zeros of the current 
and this should be performed numerically. Otherwise, the backflow 
may be also computed through a numerical integration of the 
negative part of the current $\frac{1}{2}\left(\left|\tilde{j}
\left(\tilde{t}\right)\right|-\tilde{j}\left(\tilde{t}\right)\right)$
over an interval containing the most negative peak. 
This method does not require the exact knowledge of the zeros, 
though it requires to check that only the right peak is contained 
within the integration interval.
\par
\begin{figure}[h!]
\includegraphics{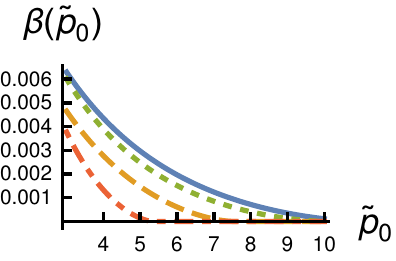}
\includegraphics{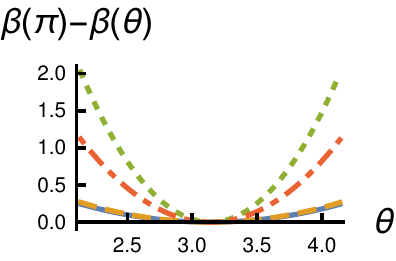}
\caption{(Color online) Backflow $\beta$ as a function of $\theta$ and $\tilde{p}_0$, 
the different curves represent a different choice of the couple of parameters $\alpha$ and $\tilde{\delta}$:
 $\alpha=2$, $\tilde{\delta}=11$(solid blue); $\alpha=3$, $\tilde{\delta}=15$ (dashed orange);
$\alpha=1.8$, $\tilde{\delta}=5$ (dotted green); $\alpha=2.5$, $\tilde{\delta}=8$ (dot-dashed red).
Left panel: $\beta$ as a function of
$\tilde{p}_{0}$, with $\theta=\pi$. Right panel: the difference between $\beta$ as a function of $\theta$ and $\beta$ obtained for $\theta = \pi$, with $\tilde{p}_{0}=3$; the values on the ordinate axis are in units of $10^{-4}$.} \label{fig:bfthetap0}
\end{figure}
\begin{figure}[h!]
\begin{center}
\includegraphics[width=8cm]{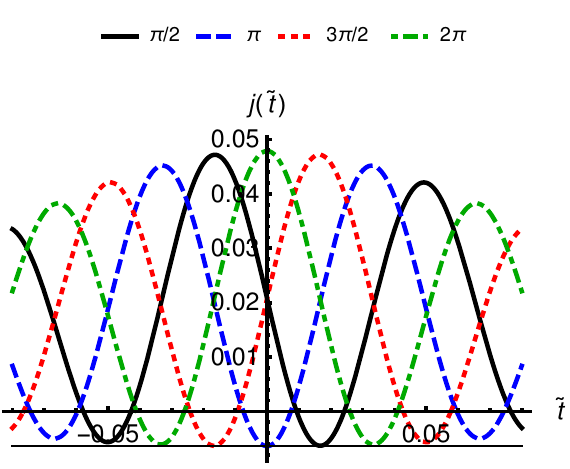}
\end{center}
\caption{(Color online) The probability current
$j\left(\tilde{t}\right)$ as a function of $\tilde{t}$ for different
values of $\theta$ at fixed values of the other parameters ($\alpha=2$,
$\tilde{\delta}=11$ and $\tilde{p}_{0}=3$). The horizontal line
highlights that the global minimum corresponds to the central negative
peak.} \label{fig: effect of theta}
\end{figure}

We now proceed to analyze the behavior of the backflow as a function 
of the different parameters. At first, we notice that the 
$\beta\left[\psi\right]$ is a decreasing function of $\tilde{p}_{0}$, 
at any fixed set of values of the other parameters, see the left panel
of Fig. \ref{fig:bfthetap0}.  Maximum backflow
is therefore attained by fixing $\tilde{p}_{0}$ to its lowest allowed
value; as mentioned above we choose $\tilde{p}_{0}=3$ to ensure a
vanishing negative momentum component.
The effect of the parameter $\theta$ is that of shifting the 
position of negative peaks of the current along the time axis, 
as it may be seen in Fig. \ref{fig: effect of theta}.
Intuition suggests that maximum backflow is obtained for a current with
a minimum located in $\tilde{t}=0$, i.e. $\theta=\pi$. Actually, the 
central peak is not always the one corresponding to the greatest backflow;
nonetheless, in order to simplify our analysis, we focus on a 
parameter range for which the central peak is the most negative one.
\par
\begin{figure}[h!]
\begin{center}
\includegraphics[width=8cm]{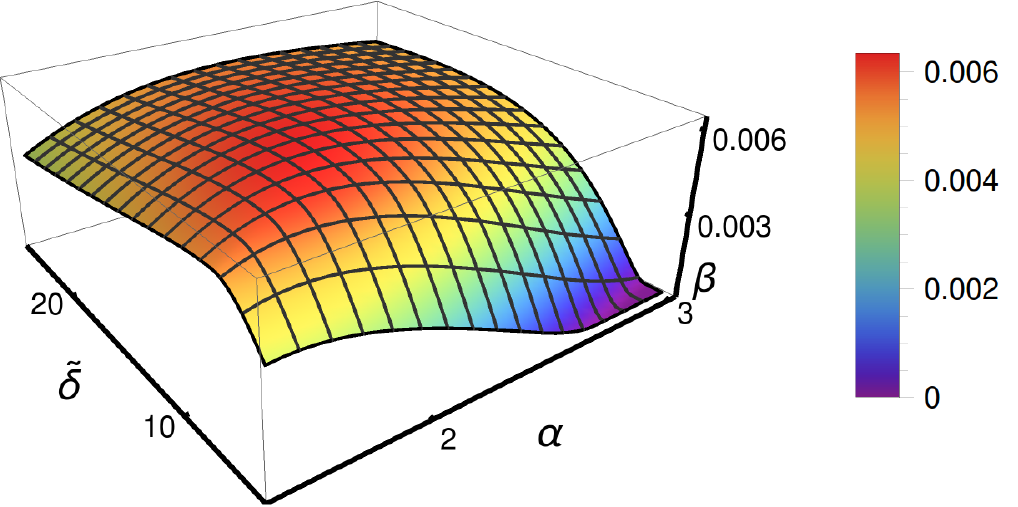}
\end{center}
\caption{(Color online) The backflow $\beta$ computed numerically as a
function of parameters $\alpha$ and $\tilde{\delta}$ in the range
$\alpha \in [1.5,3]$, $\tilde{\delta} \in [5,25]$.}
\label{fig: backflow}
\end{figure}

Unless otherwise specified, from now on we fix the values 
$\tilde{p}_{0}=3$ and $\theta=\pi$ and investigate the 
dependence of backflow on the parameters $\alpha$ and 
$\tilde{\delta}$. In particular, we explore the first-quadrant 
region of the $\left(\alpha,\tilde{\delta}\right)$ plane bounded by the lines
$\alpha=1$ and 
$\alpha=1+ \tilde{\delta}/\tilde{p}_{0}$
(which is obtained by imposing
$\tilde{j}\left(0\right)\leq0$).  For different values of $\theta$ 
other regions may be found where backflow is present, 
but no analytic expression can be found. The backflow $\beta[\psi]$
as a function of $\alpha$ and $\tilde{\delta}$ is shown in Fig. 
\ref{fig: backflow}. We can see that $\beta[\psi]$ shows a maximum, 
from which it decreases going towards the boundaries of the region.  The
maximum is obtained for $\alpha\simeq 1.9$, $\tilde{\delta}\simeq11$,
corresponding to $\beta\left[\psi\right]\simeq0.0063$ (a value slightly
larger than the one found in Ref. \cite{Yearsley2012}).  The region
closer to the value $\alpha=1$ is not shown in the plot as the backflow
is not given by the central peak.
\subsection{Quantum backflow and Wigner nonclassicality for
Gaussian superpositions} 
The Wigner function of the superposition state 
in Eq. \eqref{eq:double gaussian state} is given by 
\begin{equation}
\label{eq:wig_gaussians}
\begin{split}
& W_0(\tilde{x},\tilde{p}) =
\frac{1}{\pi \left( 1+\alpha^2+2\alpha e^{-\frac{\tilde{\delta}^2}{2}} \cos \theta \right)} \cdot \\ 
&e^{-\tilde{x}^2 /2}\Biggl[\alpha^2 e^{-2 \left(\tilde{p}-\tilde{p}_{0}\right)^2}
+e^{-2 \left(\tilde{p}-\tilde{p}_{0} - \tilde{\delta} \right)^2} 
+2\alpha\cos\left(\tilde{x}\tilde{\delta}-\theta\right) e^{-2 \left(\tilde{p}-\tilde{p}_{0}-\frac{\tilde{\delta}}{2}\right)^2}\Biggr],
\end{split}
\end{equation}
where, consistently with Eq. \eqref{eq:rescaled_parameters}, we 
used the rescaled variables
\begin{align}
\tilde{x}= \frac{x}{\sigma} \qquad \tilde{p}= \sigma p\,.
\end{align}
Notice that the rescaling is not altering the volume element and that
$W_0(\tilde{x},\tilde{p}) $ does not explicitly depend on $\sigma$.
This means that also the Wigner negativity $\Delta$, as it happens
for the backflow $\beta$, does not depend on $\sigma$.
\par
\begin{figure}[h!]
\centering
\includegraphics[width=7cm]{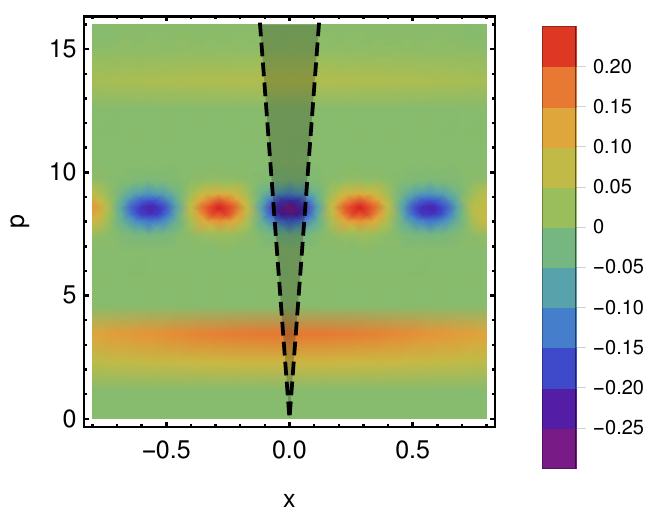}
\caption{(Color online) 
Density plot of the Wigner function of the 	superposition of Gaussian states with the
maximum backflow, the integration region $\Omega$ is the shadowed region
between the two dashed lines corresponding to $p=-\frac{m}{t_1}x$ and
$p=-\frac{m}{t_2}x$.} \label{fig:region_gaussian}
\end{figure}
\begin{figure}[h]
\centering
\includegraphics{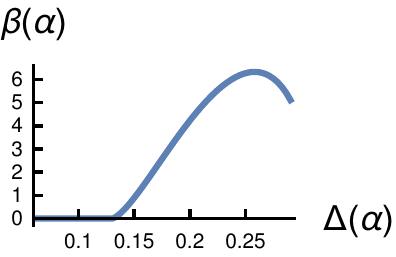}
\includegraphics{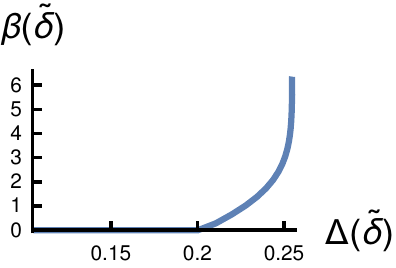}\\
\caption{(Color online) Left panel: 
parametric plot of the backflow $\beta$ as a function of the
nonclassicality $\Delta$ for $\tilde{\delta}=11$ and varying $\alpha \in
[1.5,10]$. Right panel: parametric plot of the backflow $\beta$ as a
function of the nonclassicality $\Delta$ for $\alpha=2$ and varying
$\tilde{\delta} \in [1,20]$.} \label{fig:parametri_bf_wig}
\end{figure}

The Wigner function in Eq. (\ref{eq:wig_gaussians}) is characterized 
by two Gaussian peaks corresponding to the two momenta
$p_0$ and $p_0 + \delta$ and by an interference region 
located halfway between the two peaks. In Fig. 
\ref{fig:region_gaussian} we show a contour plot of the Wigner function,
which provides an intuitive explanation to the behavior of the
backflow.
On the one hand, 
the interference effects (and thus the negative regions
of the Wigner function) are more pronounced if the the amplitude of the two
Gaussians is the same (i.e. for $\alpha = 1$).  However, this is not
leading to maximum backflow, since the Gaussian peaked at $p_0
+ \delta$ prevails in the integration region. These considerations 
also suggest that no monotonic relation between Wigner nonclassicality 
and quantum backflow may be found. As a matter of fact, 
since the positive parts of the Wigner function in the region 
$\Omega$ may compensate for the negative ones, it 
is possible to find states not showing backflow despite having negative 
Wigner function. Moreover, we may also find pairs of states with 
increasing backflow but decreasing negativity.
This non monotonic behavior of the backflow 
is illustrated in Fig. \ref{fig:parametri_bf_wig}, 
where parametric plots of the backflow as a function of the Wigner
negativity are shown for varying $\alpha$ 
or $\tilde{\delta}$.
\par
Finally, we point out that quantum backflow exhibits \emph{sudden death}
for some values of the parameters.  As for example, if $\alpha$ is
bigger than the threshold value $\alpha=1+ \tilde{\delta}/\tilde{p}_{0}$
there is no backflow. Analogue threshold values for $\tilde{\delta}$ at
fixed $\alpha$ may be found. On the contrary, Wigner negativity due to
the interference fringes dies only asymptotically, i.e. when a single
Gaussian state is recovered.  This remarkable difference may be observed
in both panels of Fig.  \ref{fig:parametri_bf_wig}, where we have
regions with no backflow but nonzero Wigner negativity.  In Appendix
\ref{app:redef_current} we present the construction of a current-like
quantity, for which the corresponding flux monotonically increases
with the Wigner negativity. However, this quantity does depend 
on the state and therefore it cannot strictly represent an 
observable.
\section{Backflow and phase space smoothing}
\label{sec:gaussian_smoothing}
We now study how robust the backflow effect is against the 
addition of thermal noise, corresponding to Gaussian 
smoothing in the phase space. We start by recalling some 
notions about  $s$-ordered quasiprobabilities, in order to 
emphasize the similarity of our analysis to the notion 
of nonclassical depth. 
\subsection{\textit{s}-ordered quasiprobability distributions}
The Wigner function can be generalized to the family of $s$-ordered
quasiprobability distributions \cite{Cahill1969a,Barnett1997}, which are
routinely used in quantum statistical optics to obtain expectation values by
averaging over the phase space.
A quasiprobability distribution $W(x,p,s)$ is labeled by 
the index $-1 \leq s \leq 1$, which reflects a particular choice 
of the ordering of the canonical operators in the expectation 
value to be computed.  For the specific values $s=1,0,-1$ we have 
the Glauber $P$ function (normal ordering), the Wigner function 
(symmetrical ordering) and the Husimi $Q$ function (antinormal
ordering), respectively. For $s<s'$, two quasiprobabilities of 
different ordering are connected through a Gaussian convolution 
\begin{equation}
\label{eq:sorderedqprob}
\begin{split}
W(x,p,s) &= W(x,p,s') \star G(x,p,s'-s) \\ 
&=  \int \mathrm{d}x' \mathrm{d}p' W(x',p',s') G(x-x',p-p',s'-s);
\end{split}
\end{equation}
where $\star$ denotes convolution and the function $G$ is a 
Gaussian defined as
\begin{equation}
G(x,p,\kappa)=\frac{1}{\pi \kappa } \exp\left[ -\frac{x^2 + p^2}{\kappa}
\right].  \end{equation}
>From Eq. \eqref{eq:sorderedqprob} one sees that going from
$s=1$ to $s=-1$ the distributions gradually become well-behaved and
positive definite functions, thanks to the Gaussian smoothing.  This is
the idea leading to the definition of 
the nonclassical depth \cite{Lee1991}, which is
a widely used method to quantify the amount of nonclassicality of a
quantum state. Basically, this quantity is the value of $s$ closer to
$s=1$ corresponding to a positive and non-singular distribution 
for a given state. In other words, the
nonclassical depth represents the minimum amount of convolution needed
in order to obtain a well-defined probability distribution from the $P$
function of a given state. For any given state, 
all the negativities in $W(x,p,s)$ must vanish for $s=-1$, 
as the $Q$ function is always non-negative by definition.
In particular, for a pure state which is not Gaussian, we have that for
all the values $-1<s\leq 0$ the quasiprobability distributions assume
negative values; i.e. for such states the $Q$ function is the only
distribution which is not negative \cite{Lutkenhaus1995}.  This means
that all non Gaussian pure states (as the superpositions of Gaussian
states we have considered so far) saturate the nonclassical depth, being
all maximally and equally nonclassical according to this criterion.
\subsection{\textit{s}-dependent current}
Here, in order to assess the robustness of backflow against noise, 
we are going to consider a generalized definition of the 
probability current based on the $s$-ordered 
quasiprobability distributions. Notice that in principle 
only the Wigner function may be used to compute the current 
via Eq. \eqref{eq: wigner flow through x=0} since the Wigner function  
is the only $s$-ordered quasiprobability distribution 
that has position and momentum probability distributions as marginals.
\footnote{Other 
generalized distributions in phase space that satisfy this property 
exist~\cite{Cohen1966a,Loughlin2003}, but they are not relevant for the
topic discussed here.}
On the other hand, introducing generalized {\it s}-dependent currents 
is meaningful if we note that the convolution of a Wigner function 
with a Gaussian represents the Wigner function of the quantum state 
after the interaction with a thermal environment.
Let us consider the master equation of a system interacting with
a bosonic bath, expressed in terms of the canonical operators 
\cite{Vacchini2002}
\begin{equation}
\begin{split}
\dot{\rho} =& - \frac{\I \gamma}{2} (2 \bar{n} + 1) \left( \left[x,\{x,\rho \}\right] - \left[p,\{x,\rho\}\right] \right) \\
&- \frac{\gamma}{2} (2\bar{n} +1) \left( [x,[x,\rho]] + [p,[p,\rho]] \right),
\end{split}
\end{equation}
where $\gamma$ is a (small) damping coefficient and 
$\bar{n}$ is the average photon number of the thermal environment. 
In terms of the Wigner function, the solution of the above equation
may be written as 
\begin{equation}
\label{eq:ME_sol_phase_space}
e^{-2\tau} W_t(e^{-\tau} x,e^{-\tau}p)=W_0(x,p) \star G(x,p,-s_\tau),
\end{equation}
where $\tau = \gamma t$ and $s_{\tau}=-2(2 \bar{n} + 1)
(e^{2\tau} -1)$, see e.g. \cite{Barnett1997} for details.
$W(x,p,s)$ is thus the Wigner function of the state obtained from 
the initial one after the interaction with a noisy environment.
Notice that the rescaling due to dissipation, i.e. the exponential of $\tau$ 
appearing on the l.h.s. of \eqref{eq:ME_sol_phase_space}, plays no role 
in determining the negativity of the Wigner function and 
the backflow.

If the initial state has the Wigner function $W_0(x,p)$, 
the state after the noisy interaction has a Wigner function
given by 
\begin{equation}
\label{eq:s ordered initial wigner}
W_0(x,p,s) = W_0(x,p) \star G(x,p,-s),
\end{equation}
where $s$ will in general be a function of the temperature, of the damping coefficient
and of the interaction time. At this point, we consider 
$W_0(x,p,s)$ as the \emph{initial} Wigner function of a mixed 
state evolving according to the free particle Hamiltonian,
and we get an $s$-dependent and time dependent Wigner function
\begin{equation}
W_t(x,p,s)=W_0\left( x-\frac{p}{m} t, p,s \right),
\end{equation} 
which can in turn be used to compute the $s$-dependent current 
\begin{equation}
\label{eq:s ordered current}
j(t,s) = \int_{-\infty}^{+\infty} \mathrm{d} p \frac{p}{m} W_t(0,p,s).
\end{equation}
We only consider $-1\leq s \leq 0$, in order to have a 
smoothing of the initial Wigner function; in terms of ordering 
this means going from the Wigner towards the $Q$ function.
\par
One may wonder what happens if we exchange
the order of the evolution in time and the convolution. This means 
convolving the time-dependent Wigner function $W_t(x,p)$,
instead of convolving the initial one and then applying the free 
evolution. This way of proceeding is conceptually different 
and indeed yields slightly different numerical results,
but the qualitative behavior is unchanged. 
Before proceeding we also notice that our scheme is different 
from considering the backflow of an open quantum system, where
the expression for the probability current may be different 
\cite{Yearsley2010}.
\subsection{\textit{s}-dependent backflow and negative current depth}
\begin{figure}[h!]
\includegraphics{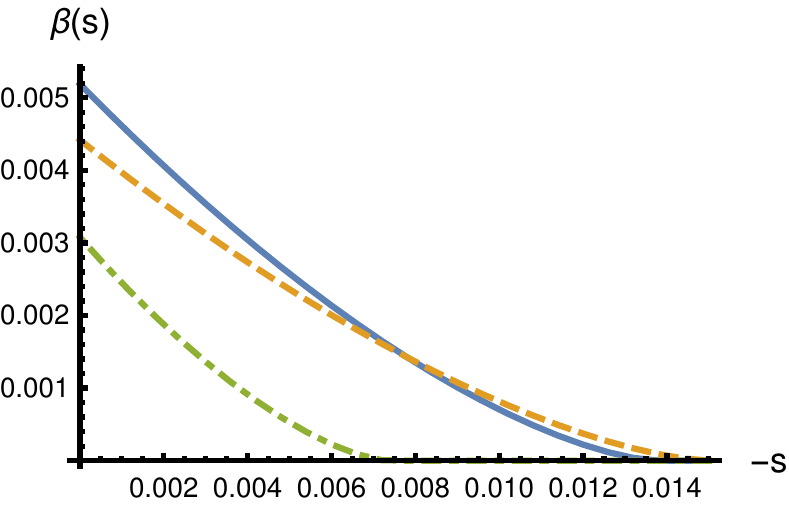}
\caption{(Color online) Plots of the $s$-dependent backflow as a function of the
Gaussian smoothing parameter $s$. From top to bottom in the region
$s\approx0$ we have the states with $\alpha=2$ and $\tilde{\delta}=7$
(solid blue), $\alpha=2$ and $\tilde{\delta}=6$ (dashed orange),
$\alpha=3$ and $\tilde{\delta}=10$ (dot-dashed green).}
\label{fig:s_dep_bf}
\end{figure}

Having defined an $s$-dependent current we can straightforwardly apply
the definition of backflow \eqref{eq:flux function} and obtain an
$s$-dependent backflow.  As we can see in Fig. \ref{fig:s_dep_bf} the
backflow vanishes for a certain $s > -1$, i.e. it exhibits sudden death, 
in contrast with the negativity of the Wigner function.
Having more backflow initially (for $s=0$) usually means that the
backflow of the state survives longer (i.e. it disappears for a value of
$s$ closer to $-1$).  However, as it may be seen from Fig. 
\ref{fig:s_dep_bf}, this is not necessarily the case for any choice 
of the parameters.
\par
In order to better analyze this behavior we introduce, in analogy
with the nonclassical depth, the \emph{negative current depth}, which
is defined as follows. Upon denoting by $C$ the subinterval of
$s \in [-1,0]$ leading to a positive $s$-dependent current 
in \eqref{eq:s ordered current},
then the negative current depth $s_m$ is defined as 
\begin{equation}
s_m \equiv \inf_{s\in C} (-s),
\end{equation} 
which is a positive quantity bounded between $0$ and $1$.
This quantity provide an alternative quantification of backflow;
instead of quantifying how much probability is flowing 
backwards we quantify the amount of Gaussian convolution, i.e. noise,
needed to destroy the backflow effect of a given initial state.
\begin{figure}[h]
\centering
\includegraphics{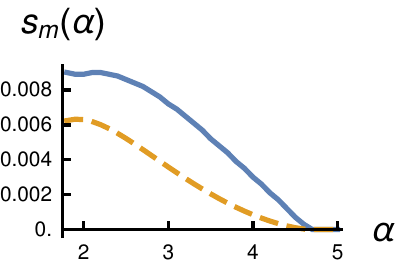}
\includegraphics{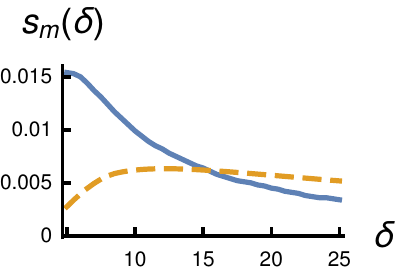}\\
\caption{(Color online) The negative current depth (solid blue) and the 
backflow (dashed orange). Left panel: the quantities are shown as a 
function of $\alpha$ for $\tilde{\delta}=11$. Right panel: as a function 
of $\tilde{\delta}$ for $\alpha=2$.}
\label{fig:negcurrdepth}
\end{figure}
\par
Fig. \ref{fig:negcurrdepth} shows that the negative current 
depth and the backflow of a quantum state
have similar behavior, but regions where they are not 
monotonic exist, as it can be seen in the right panel.
Upon looking at the values of $s_m$ in Fig. \ref{fig:negcurrdepth}, 
and since for all these states the negativities of the Wigner 
function completely disappear only for $s=-1$, we conclude 
that the backflow is a very fragile form of nonclassicality.
Notice that other criteria for nonclassicality exist and their 
behavior for superpositions of Gaussian states in the presence 
of a thermal environment has been studied \cite{Paavola2011a}.
Results have shown that almost all these indicators vanish for a 
finite Gaussian smoothing, except the visibility of the 
interference fringes which vanish only asymptotically.
\section{Conclusions}
\label{sec:out}
The quantum backflow effect is a counterintuitive behavior observed 
in the probability current of a free particle, which may be negative even 
for states with vanishing negative momentum component. Quantum backflow
may be described in the phase space, showing that its occurrence 
is connected to the {\em classical phase space dynamics} of
a {\em nonclassical initial state}.  In the case of the free particle such
flow in phase space is directly connected to the probability of
observing the particle in a certain region, thus it is relevant in
physical problems such as determining the time of arrival.  The reason
for this counterintuitive behavior lies in the fact that a state with
only positive momenta does not need to have a positive probability
current. 
Since the backflow effect, despite being dynamical, is completely due to the
nonclassical character of the initial state, we investigated how this
kind of nonclassicality compares to 
the negativity of the Wigner function.
In order to carry out this investigation we have focused attention 
to superpositions of two Gaussian states.
\par
We found that the two notions of nonclassicality are inequivalent and
the respective quantifiers do not show a monotonic behavior.
We have then further characterized the backflow effect, by studying its
resilience to the operation of Gaussian smoothing in phase space, which
describes the interaction of the initial state with a thermal environment.
\par
Overall, our results suggest that backflow is connected to 
a more restrictive notion of nonclassicality, the negativity 
of the Wigner function being just a necessary condition for its
occurrence. Backflow has a different behavior in terms of the
defining parameters of the state, in particular it vanishes for some
threshold values. Moreover, the negativity of the probability current is
a feature which is more easily destroyed by a thermal environment than
the negativity of the Wigner function itself, confirming the idea that
backflow is a nonclassical effect of a higher order and thus more fragile.

\appendix
\section{A redefined current}
\label{app:redef_current}
Having concluded in Section \ref{sec:gaussians} that backflow is not
monotonically linked to the negative Wigner function volumes, one may
try to understand if these negative volumes can actually be linked to
some current-like quantity. To this aim, one would have to restrict
integral (\ref{eq: wigner flow through x=0}) from the whole $x=0$
axis, to the sole parts where it is crossed by the negative volumes in
their phase space motion. This restriction gives the following expression:
\begin{equation}
\label{eq:integral only neg volume of W}
\begin{split}
&\int_{p_{1}}^{p_{2}} \! \mathrm{d}p\frac{p}{m} \, W_{t}\left(0,p\right)= \\
 = \int_{-\infty}^{+\infty}  \! \mathrm{d}x \, &\left(\int_{p_{1}}^{p_{2}}\frac{1}{2\pi}e^{ipx}\mathrm{d}p\right)\frac{-i}{2m}\Bigg[\psi^{\ast}\left(\frac{x}{2},t\right)\frac{\partial}{\partial x}\psi\left(-\frac{x}{2},t\right)\\
&+\psi\left(-\frac{x}{2},t\right)\frac{\partial}{\partial x}\psi^{\ast}\left(-\frac{x}{2},t\right)\Bigg].
\end{split}
\end{equation}
A closer examination of this expression induces us to consider a new quantity defined as
\begin{equation}
\label{eq:eta}
\eta_{p_{1},p_{2}}\left(t\right):=\int_{-\infty}^{+\infty}
\!\!\!
\mathrm{d}x\,\delta_{p_{1},p_{2}}\left(x\right)\, J(x,t)
\end{equation}
where
$$
J(x,t)=
\frac{i}{2m}\left(\psi\left(x,t\right)\frac{\partial}{\partial
x}\psi^{*}\left(x,t\right)-\psi^{*}\left(x,t\right)\frac{\partial}{\partial
x}\psi\left(x,t\right) \right)
$$
and
\begin{equation}
\delta_{p_{1},p_{2}}\left(x\right):=\frac{1}{\pi x}\sin\left(p_{1}x\right)-\frac{1}{\pi x}\sin\left(p_{2}x\right).
\end{equation}
This quantity (defined by parameters $p_{1}$ and $p_{2}$ which have the dimensions of
momentum) has a similar structure to \eqref{eq:integral only neg volume of W} but takes
real values for all times $t$. Accordingly we expect its negative flux over time to behave
in a similar way to the negative volume of the Wigner function.

To give a physical interpretation of this newly defined $\eta_{p_{1},p_{2}}$, it is interesting
to observe that it is the difference of two objects, each one expressed as the convolution of the
ordinary current $J\left(x,t\right)$ with a smooth approximation of the Dirac delta. Such
objects can be thought to represent a probability current arising from a realistic position measurement.
These measurements have a finite spatial resolution given by the widths $\frac{1}{p_{1}}$ and
$\frac{1}{p_{2}}$ of the approximations of the delta used. This leads us to interpret
$\eta_{p_{1},p_{2}}$ as the difference between two realistic current measurements
with different spatial resolutions given by the inverses of $p_{1}$ and $p_{2}$.

\begin{figure}[!htbp]
	\centering
	\includegraphics{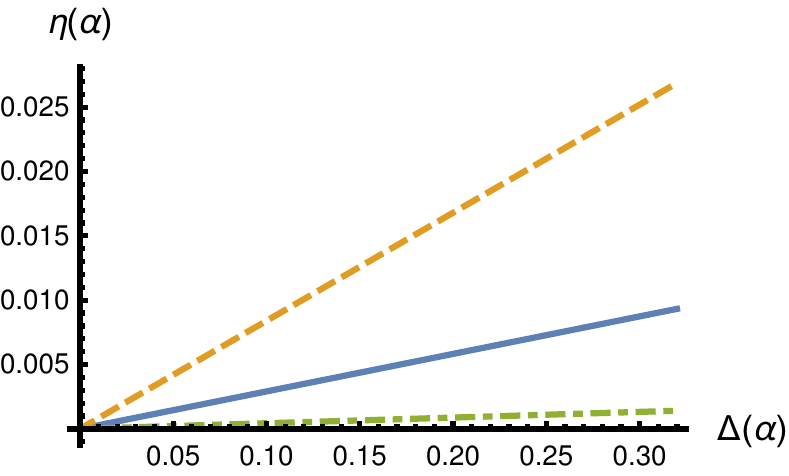}
	\caption{ (Color online) Parametric plot of the negative flow of $\eta_{p_{1},p_{2}}$ versus the nonclassicality $\Delta$ for $\tilde{\delta}=9.5$ (dot-dashed green), $\tilde{\delta}=10$ (solid blue) and $\tilde{\delta}=10.5$ (dashed orange). The parameter $\alpha$ varies in the range $\alpha\in\left[0.01,5\right]$, while the two defining parameters of the new current are kept fixed at the values $p_{1}=7$ and $p_{2}=9$.}
	\label{fig:parametric eta}
\end{figure}

Of course the construction of this currents depends strongly on the values of parameters
$p_{1}$ and $p_{2}$, which have to be chosen in such a way as to ensure that the negative
volume of the state considered passes through the interval $\left(p_{1},p_{2}\right)$ in its motion
through phase space. As this choice is state dependent, the flux of $\eta_{p_{1},p_{2}}$ cannot
be considered a true observable. However, for the superpositions of Gaussians previously used,
with a fixed value of $\tilde{\delta}$ and for consequently chosen values of $p_{1}$ and $p_{2}$,
the negative flux of $\eta_{p_{1},p_{2}}\left(t\right)$ shows a dependence on parameter
$\alpha$ which is remarkably similar to that of the total negative volume of the Wigner function $\Delta$.
This is well illustrated by Fig. \ref{fig:parametric eta}, which shows the parametric dependence
of the negative flux of $\eta_{p_{1},p_{2}}$ on nonclassicality $\Delta$ for varying values of
the parameter $\alpha$ and at fixed values of $\tilde{\delta}$. Equivalent results are found for
different values of $\tilde{\delta}$.

\begin{acknowledgments}
This work has been supported by EU through the
Collaborative Project QuProCS (Grant Agreement 641277) and by UniMI
through the H2020 Transition Grant 15-6-3008000-625.
\end{acknowledgments}

\bibliography{qbkfl7}

\begin{thebibliography}{58}%
\makeatletter
\providecommand \@ifxundefined [1]{%
 \@ifx{#1\undefined}
}%
\providecommand \@ifnum [1]{%
 \ifnum #1\expandafter \@firstoftwo
 \else \expandafter \@secondoftwo
 \fi
}%
\providecommand \@ifx [1]{%
 \ifx #1\expandafter \@firstoftwo
 \else \expandafter \@secondoftwo
 \fi
}%
\providecommand \natexlab [1]{#1}%
\providecommand \enquote  [1]{``#1''}%
\providecommand \bibnamefont  [1]{#1}%
\providecommand \bibfnamefont [1]{#1}%
\providecommand \citenamefont [1]{#1}%
\providecommand \href@noop [0]{\@secondoftwo}%
\providecommand \href [0]{\begingroup \@sanitize@url \@href}%
\providecommand \@href[1]{\@@startlink{#1}\@@href}%
\providecommand \@@href[1]{\endgroup#1\@@endlink}%
\providecommand \@sanitize@url [0]{\catcode `\\12\catcode `\$12\catcode
  `\&12\catcode `\#12\catcode `\^12\catcode `\_12\catcode `\%12\relax}%
\providecommand \@@startlink[1]{}%
\providecommand \@@endlink[0]{}%
\providecommand \url  [0]{\begingroup\@sanitize@url \@url }%
\providecommand \@url [1]{\endgroup\@href {#1}{\urlprefix }}%
\providecommand \urlprefix  [0]{URL }%
\providecommand \Eprint [0]{\href }%
\providecommand \doibase [0]{http://dx.doi.org/}%
\providecommand \selectlanguage [0]{\@gobble}%
\providecommand \bibinfo  [0]{\@secondoftwo}%
\providecommand \bibfield  [0]{\@secondoftwo}%
\providecommand \translation [1]{[#1]}%
\providecommand \BibitemOpen [0]{}%
\providecommand \bibitemStop [0]{}%
\providecommand \bibitemNoStop [0]{.\EOS\space}%
\providecommand \EOS [0]{\spacefactor3000\relax}%
\providecommand \BibitemShut  [1]{\csname bibitem#1\endcsname}%
\let\auto@bib@innerbib\@empty
\bibitem [{\citenamefont {Allcock}(1969)}]{Allcock1969}%
  \BibitemOpen
  \bibfield  {author} {\bibinfo {author} {\bibfnamefont {G.}~\bibnamefont
  {Allcock}},\ }\href {\doibase http://dx.doi.org/10.1016/0003-4916(69)90251-6}
  {\bibfield  {journal} {\bibinfo  {journal} {Ann. Phys.}\ }\textbf {\bibinfo
  {volume} {53}},\ \bibinfo {pages} {253 } (\bibinfo {year}
  {1969})}\BibitemShut {NoStop}%
\bibitem [{\citenamefont {Bracken}\ and\ \citenamefont
  {Melloy}(1994)}]{Bracken1994}%
  \BibitemOpen
  \bibfield  {author} {\bibinfo {author} {\bibfnamefont {A.~J.}\ \bibnamefont
  {Bracken}}\ and\ \bibinfo {author} {\bibfnamefont {G.~F.}\ \bibnamefont
  {Melloy}},\ }\href {\doibase 10.1088/0305-4470/27/6/040} {\bibfield
  {journal} {\bibinfo  {journal} {J. Phys. A}\ }\textbf {\bibinfo {volume}
  {27}},\ \bibinfo {pages} {2197} (\bibinfo {year} {1994})}\BibitemShut
  {NoStop}%
\bibitem [{\citenamefont {Berry}(2010)}]{mvb10}%
  \BibitemOpen
  \bibfield  {author} {\bibinfo {author} {\bibfnamefont {M.~V.}\ \bibnamefont
  {Berry}},\ }\href@noop {} {\bibfield  {journal} {\bibinfo  {journal} {J.
  Phys. A}\ }\textbf {\bibinfo {volume} {43}},\ \bibinfo {pages} {415302}
  (\bibinfo {year} {2010})}\BibitemShut {NoStop}%
\bibitem [{\citenamefont {Strange}(2012{\natexlab{a}})}]{stra12}%
  \BibitemOpen
  \bibfield  {author} {\bibinfo {author} {\bibfnamefont {P.}~\bibnamefont
  {Strange}},\ }\href@noop {} {\bibfield  {journal} {\bibinfo  {journal} {Eur.
  J. Phys.}\ }\textbf {\bibinfo {volume} {33}},\ \bibinfo {pages} {1147}
  (\bibinfo {year} {2012}{\natexlab{a}})}\BibitemShut {NoStop}%
\bibitem [{\citenamefont {Eveson}\ \emph {et~al.}(2005)\citenamefont {Eveson},
  \citenamefont {Fewster},\ and\ \citenamefont {Verch}}]{Eveson2005}%
  \BibitemOpen
  \bibfield  {author} {\bibinfo {author} {\bibfnamefont {S.~P.}\ \bibnamefont
  {Eveson}}, \bibinfo {author} {\bibfnamefont {C.~J.}\ \bibnamefont {Fewster}},
  \ and\ \bibinfo {author} {\bibfnamefont {R.}~\bibnamefont {Verch}},\ }\href
  {\doibase 10.1007/s00023-005-0197-9} {\bibfield  {journal} {\bibinfo
  {journal} {Ann. Henri Poincar{\'{e}}}\ }\textbf {\bibinfo {volume} {6}},\
  \bibinfo {pages} {1} (\bibinfo {year} {2005})}\BibitemShut {NoStop}%
\bibitem [{\citenamefont {Penz}\ \emph {et~al.}(2006)\citenamefont {Penz},
  \citenamefont {Gr{\"{u}}bl}, \citenamefont {Kreidl},\ and\ \citenamefont
  {Wagner}}]{Penz2006a}%
  \BibitemOpen
  \bibfield  {author} {\bibinfo {author} {\bibfnamefont {M.}~\bibnamefont
  {Penz}}, \bibinfo {author} {\bibfnamefont {G.}~\bibnamefont {Gr{\"{u}}bl}},
  \bibinfo {author} {\bibfnamefont {S.}~\bibnamefont {Kreidl}}, \ and\ \bibinfo
  {author} {\bibfnamefont {P.}~\bibnamefont {Wagner}},\ }\href {\doibase
  10.1088/0305-4470/39/2/012} {\bibfield  {journal} {\bibinfo  {journal} {J.
  Phys. A}\ }\textbf {\bibinfo {volume} {39}},\ \bibinfo {pages} {423}
  (\bibinfo {year} {2006})}\BibitemShut {NoStop}%
\bibitem [{\citenamefont {Muga}\ \emph {et~al.}(1999)\citenamefont {Muga},
  \citenamefont {Palao},\ and\ \citenamefont {Leavens}}]{Muga1999}%
  \BibitemOpen
  \bibfield  {author} {\bibinfo {author} {\bibfnamefont {J.~G.}\ \bibnamefont
  {Muga}}, \bibinfo {author} {\bibfnamefont {J.~P.}\ \bibnamefont {Palao}}, \
  and\ \bibinfo {author} {\bibfnamefont {C.}~\bibnamefont {Leavens}},\ }\href
  {\doibase 10.1016/S0375-9601(99)00020-1} {\bibfield  {journal} {\bibinfo
  {journal} {Phys. Lett. A}\ }\textbf {\bibinfo {volume} {253}},\ \bibinfo
  {pages} {21} (\bibinfo {year} {1999})}\BibitemShut {NoStop}%
\bibitem [{\citenamefont {Yearsley}\ \emph {et~al.}(2012)\citenamefont
  {Yearsley}, \citenamefont {Halliwell}, \citenamefont {Hartshorn},\ and\
  \citenamefont {Whitby}}]{Yearsley2012}%
  \BibitemOpen
  \bibfield  {author} {\bibinfo {author} {\bibfnamefont {J.~M.}\ \bibnamefont
  {Yearsley}}, \bibinfo {author} {\bibfnamefont {J.~J.}\ \bibnamefont
  {Halliwell}}, \bibinfo {author} {\bibfnamefont {R.}~\bibnamefont
  {Hartshorn}}, \ and\ \bibinfo {author} {\bibfnamefont {A.}~\bibnamefont
  {Whitby}},\ }\href {\doibase 10.1103/PhysRevA.86.042116} {\bibfield
  {journal} {\bibinfo  {journal} {Phys. Rev. A}\ }\textbf {\bibinfo {volume}
  {86}},\ \bibinfo {pages} {042116} (\bibinfo {year} {2012})}\BibitemShut
  {NoStop}%
\bibitem [{\citenamefont {Halliwell}\ \emph {et~al.}(2013)\citenamefont
  {Halliwell}, \citenamefont {Gillman}, \citenamefont {Lennon}, \citenamefont
  {Patel},\ and\ \citenamefont {Ramirez}}]{Halliwell2013}%
  \BibitemOpen
  \bibfield  {author} {\bibinfo {author} {\bibfnamefont {J.~J.}\ \bibnamefont
  {Halliwell}}, \bibinfo {author} {\bibfnamefont {E.}~\bibnamefont {Gillman}},
  \bibinfo {author} {\bibfnamefont {O.}~\bibnamefont {Lennon}}, \bibinfo
  {author} {\bibfnamefont {M.}~\bibnamefont {Patel}}, \ and\ \bibinfo {author}
  {\bibfnamefont {I.}~\bibnamefont {Ramirez}},\ }\href {\doibase
  10.1088/1751-8113/46/47/475303} {\bibfield  {journal} {\bibinfo  {journal}
  {J. Phys. A}\ }\textbf {\bibinfo {volume} {46}},\ \bibinfo {pages} {475303}
  (\bibinfo {year} {2013})}\BibitemShut {NoStop}%
\bibitem [{\citenamefont {Palmero}\ \emph {et~al.}(2013)\citenamefont
  {Palmero}, \citenamefont {Torrontegui}, \citenamefont {Muga},\ and\
  \citenamefont {Modugno}}]{Palmero2013}%
  \BibitemOpen
  \bibfield  {author} {\bibinfo {author} {\bibfnamefont {M.}~\bibnamefont
  {Palmero}}, \bibinfo {author} {\bibfnamefont {E.}~\bibnamefont
  {Torrontegui}}, \bibinfo {author} {\bibfnamefont {J.~G.}\ \bibnamefont
  {Muga}}, \ and\ \bibinfo {author} {\bibfnamefont {M.}~\bibnamefont
  {Modugno}},\ }\href {\doibase 10.1103/PhysRevA.87.053618} {\bibfield
  {journal} {\bibinfo  {journal} {Phys. Rev. A}\ }\textbf {\bibinfo {volume}
  {87}},\ \bibinfo {pages} {053618} (\bibinfo {year} {2013})}\BibitemShut
  {NoStop}%
\bibitem [{\citenamefont {Melloy}\ and\ \citenamefont
  {Bracken}(1998{\natexlab{a}})}]{Melloy1998a}%
  \BibitemOpen
  \bibfield  {author} {\bibinfo {author} {\bibfnamefont {G.~F.}\ \bibnamefont
  {Melloy}}\ and\ \bibinfo {author} {\bibfnamefont {A.~J.}\ \bibnamefont
  {Bracken}},\ }\href {\doibase 10.1023/A:1018724313788} {\bibfield  {journal}
  {\bibinfo  {journal} {Found. Phys.}\ ,\ \bibinfo {pages} {505}} (\bibinfo
  {year} {1998}{\natexlab{a}})}\BibitemShut {NoStop}%
\bibitem [{\citenamefont {Melloy}\ and\ \citenamefont
  {Bracken}(1998{\natexlab{b}})}]{Melloy1998}%
  \BibitemOpen
  \bibfield  {author} {\bibinfo {author} {\bibfnamefont {G.~F.}\ \bibnamefont
  {Melloy}}\ and\ \bibinfo {author} {\bibfnamefont {A.~J.}\ \bibnamefont
  {Bracken}},\ }\href {\doibase
  10.1002/(SICI)1521-3889(199812)7:7/8<726::AID-ANDP726>3.0.CO;2-P} {\bibfield
  {journal} {\bibinfo  {journal} {Ann. Phys.}\ }\textbf {\bibinfo {volume}
  {7}},\ \bibinfo {pages} {726} (\bibinfo {year}
  {1998}{\natexlab{b}})}\BibitemShut {NoStop}%
\bibitem [{\citenamefont {Strange}(2012{\natexlab{b}})}]{Strange2012}%
  \BibitemOpen
  \bibfield  {author} {\bibinfo {author} {\bibfnamefont {P.}~\bibnamefont
  {Strange}},\ }\href {\doibase 10.1088/0143-0807/33/5/1147} {\bibfield
  {journal} {\bibinfo  {journal} {Eur. J. Phys.}\ }\textbf {\bibinfo {volume}
  {33}},\ \bibinfo {pages} {1147} (\bibinfo {year}
  {2012}{\natexlab{b}})}\BibitemShut {NoStop}%
\bibitem [{\citenamefont {Steuernagel}\ \emph {et~al.}(2013)\citenamefont
  {Steuernagel}, \citenamefont {Kakofengitis},\ and\ \citenamefont
  {Ritter}}]{ole13}%
  \BibitemOpen
  \bibfield  {author} {\bibinfo {author} {\bibfnamefont {O.}~\bibnamefont
  {Steuernagel}}, \bibinfo {author} {\bibfnamefont {D.}~\bibnamefont
  {Kakofengitis}}, \ and\ \bibinfo {author} {\bibfnamefont {G.}~\bibnamefont
  {Ritter}},\ }\href {\doibase 10.1103/PhysRevLett.110.030401} {\bibfield
  {journal} {\bibinfo  {journal} {Phys. Rev. Lett.}\ }\textbf {\bibinfo
  {volume} {110}},\ \bibinfo {pages} {030401} (\bibinfo {year}
  {2013})}\BibitemShut {NoStop}%
\bibitem [{\citenamefont {Wigner}(1932)}]{Wigner1932}%
  \BibitemOpen
  \bibfield  {author} {\bibinfo {author} {\bibfnamefont {E.}~\bibnamefont
  {Wigner}},\ }\href {\doibase 10.1103/PhysRev.40.749} {\bibfield  {journal}
  {\bibinfo  {journal} {Phys. Rev.}\ }\textbf {\bibinfo {volume} {40}},\
  \bibinfo {pages} {749} (\bibinfo {year} {1932})}\BibitemShut {NoStop}%
\bibitem [{\citenamefont {Hillery}\ \emph {et~al.}(1984)\citenamefont
  {Hillery}, \citenamefont {O'Connell}, \citenamefont {Scully},\ and\
  \citenamefont {Wigner}}]{Hillery1984}%
  \BibitemOpen
  \bibfield  {author} {\bibinfo {author} {\bibfnamefont {M.}~\bibnamefont
  {Hillery}}, \bibinfo {author} {\bibfnamefont {R.~F.}\ \bibnamefont
  {O'Connell}}, \bibinfo {author} {\bibfnamefont {M.}~\bibnamefont {Scully}}, \
  and\ \bibinfo {author} {\bibfnamefont {E.}~\bibnamefont {Wigner}},\ }\href
  {\doibase 10.1016/0370-1573(84)90160-1} {\bibfield  {journal} {\bibinfo
  {journal} {Phys. Rep.}\ }\textbf {\bibinfo {volume} {106}},\ \bibinfo {pages}
  {121} (\bibinfo {year} {1984})}\BibitemShut {NoStop}%
\bibitem [{\citenamefont {Feynman}(1987)}]{Feynman1987}%
  \BibitemOpen
  \bibfield  {author} {\bibinfo {author} {\bibfnamefont {R.~P.}\ \bibnamefont
  {Feynman}},\ }in\ \href@noop {} {\emph {\bibinfo {booktitle} {Quantum
  Implications: Essays in Honour of David Bohm}}},\ \bibinfo {editor} {edited
  by\ \bibinfo {editor} {\bibfnamefont {B.}~\bibnamefont {Hiley}}\ and\
  \bibinfo {editor} {\bibfnamefont {F.~D.}\ \bibnamefont {Peat}}}\ (\bibinfo
  {publisher} {Methuen},\ \bibinfo {address} {London},\ \bibinfo {year}
  {1987})\ Chap.~\bibinfo {chapter} {13}, pp.\ \bibinfo {pages}
  {235--248}\BibitemShut {NoStop}%
\bibitem [{\citenamefont {Scully}\ \emph {et~al.}(1994)\citenamefont {Scully},
  \citenamefont {Walther},\ and\ \citenamefont {Schleich}}]{Scully1994}%
  \BibitemOpen
  \bibfield  {author} {\bibinfo {author} {\bibfnamefont {M.~O.}\ \bibnamefont
  {Scully}}, \bibinfo {author} {\bibfnamefont {H.}~\bibnamefont {Walther}}, \
  and\ \bibinfo {author} {\bibfnamefont {W.~P.}\ \bibnamefont {Schleich}},\
  }\href {\doibase 10.1103/PhysRevA.49.1562} {\bibfield  {journal} {\bibinfo
  {journal} {Phys. Rev. A}\ }\textbf {\bibinfo {volume} {49}},\ \bibinfo
  {pages} {1562} (\bibinfo {year} {1994})}\BibitemShut {NoStop}%
\bibitem [{\citenamefont {Cahill}\ and\ \citenamefont
  {Glauber}(1969)}]{Cahill1969a}%
  \BibitemOpen
  \bibfield  {author} {\bibinfo {author} {\bibfnamefont {K.~E.}\ \bibnamefont
  {Cahill}}\ and\ \bibinfo {author} {\bibfnamefont {R.~J.}\ \bibnamefont
  {Glauber}},\ }\href {\doibase 10.1103/PhysRev.177.1882} {\bibfield  {journal}
  {\bibinfo  {journal} {Phys. Rev.}\ }\textbf {\bibinfo {volume} {177}},\
  \bibinfo {pages} {1882} (\bibinfo {year} {1969})}\BibitemShut {NoStop}%
\bibitem [{\citenamefont {Lee}(1991)}]{Lee1991}%
  \BibitemOpen
  \bibfield  {author} {\bibinfo {author} {\bibfnamefont {C.~T.}\ \bibnamefont
  {Lee}},\ }\href {\doibase 10.1103/PhysRevA.44.R2775} {\bibfield  {journal}
  {\bibinfo  {journal} {Phys. Rev. A}\ }\textbf {\bibinfo {volume} {44}},\
  \bibinfo {pages} {R2775} (\bibinfo {year} {1991})}\BibitemShut {NoStop}%
\bibitem [{\citenamefont {Lee}(1992)}]{Lee1992}%
  \BibitemOpen
  \bibfield  {author} {\bibinfo {author} {\bibfnamefont {C.~T.}\ \bibnamefont
  {Lee}},\ }\href {\doibase 10.1103/PhysRevA.45.6586} {\bibfield  {journal}
  {\bibinfo  {journal} {Phys. Rev. A}\ }\textbf {\bibinfo {volume} {45}},\
  \bibinfo {pages} {6586} (\bibinfo {year} {1992})}\BibitemShut {NoStop}%
\bibitem [{\citenamefont {Lee}(1995)}]{Lee1995}%
  \BibitemOpen
  \bibfield  {author} {\bibinfo {author} {\bibfnamefont {C.~T.}\ \bibnamefont
  {Lee}},\ }\href {\doibase 10.1103/PhysRevA.52.3374} {\bibfield  {journal}
  {\bibinfo  {journal} {Phys. Rev. A}\ }\textbf {\bibinfo {volume} {52}},\
  \bibinfo {pages} {3374} (\bibinfo {year} {1995})}\BibitemShut {NoStop}%
\bibitem [{\citenamefont {Dodonov}(2002)}]{dod02}%
  \BibitemOpen
  \bibfield  {author} {\bibinfo {author} {\bibfnamefont {V.~V.}\ \bibnamefont
  {Dodonov}},\ }\href {\doibase 10.1088/1464-4266/4/1/201} {\bibfield
  {journal} {\bibinfo  {journal} {J. Opt. B}\ }\textbf {\bibinfo {volume}
  {4}},\ \bibinfo {pages} {R1} (\bibinfo {year} {2002})}\BibitemShut {NoStop}%
\bibitem [{\citenamefont {Richter}\ and\ \citenamefont
  {Vogel}(2002)}]{Vogel02}%
  \BibitemOpen
  \bibfield  {author} {\bibinfo {author} {\bibfnamefont {T.}~\bibnamefont
  {Richter}}\ and\ \bibinfo {author} {\bibfnamefont {W.}~\bibnamefont
  {Vogel}},\ }\href {\doibase 10.1103/PhysRevLett.89.283601} {\bibfield
  {journal} {\bibinfo  {journal} {Phys. Rev. Lett.}\ }\textbf {\bibinfo
  {volume} {89}},\ \bibinfo {pages} {283601} (\bibinfo {year}
  {2002})}\BibitemShut {NoStop}%
\bibitem [{\citenamefont {Kim}\ \emph {et~al.}(2002)\citenamefont {Kim},
  \citenamefont {Son}, \citenamefont {Bu\ifmmode~\check{z}\else \v{z}\fi{}ek},\
  and\ \citenamefont {Knight}}]{kim02}%
  \BibitemOpen
  \bibfield  {author} {\bibinfo {author} {\bibfnamefont {M.~S.}\ \bibnamefont
  {Kim}}, \bibinfo {author} {\bibfnamefont {W.}~\bibnamefont {Son}}, \bibinfo
  {author} {\bibfnamefont {V.}~\bibnamefont {Bu\ifmmode~\check{z}\else
  \v{z}\fi{}ek}}, \ and\ \bibinfo {author} {\bibfnamefont {P.~L.}\ \bibnamefont
  {Knight}},\ }\href {\doibase 10.1103/PhysRevA.65.032323} {\bibfield
  {journal} {\bibinfo  {journal} {Phys. Rev. A}\ }\textbf {\bibinfo {volume}
  {65}},\ \bibinfo {pages} {032323} (\bibinfo {year} {2002})}\BibitemShut
  {NoStop}%
\bibitem [{\citenamefont {Xiang-bin}(2002)}]{wang02}%
  \BibitemOpen
  \bibfield  {author} {\bibinfo {author} {\bibfnamefont {W.}~\bibnamefont
  {Xiang-bin}},\ }\href {\doibase 10.1103/PhysRevA.66.024303} {\bibfield
  {journal} {\bibinfo  {journal} {Phys. Rev. A}\ }\textbf {\bibinfo {volume}
  {66}},\ \bibinfo {pages} {024303} (\bibinfo {year} {2002})}\BibitemShut
  {NoStop}%
\bibitem [{\citenamefont {Shchukin}\ \emph {et~al.}(2005)\citenamefont
  {Shchukin}, \citenamefont {Richter},\ and\ \citenamefont {Vogel}}]{Vogel05}%
  \BibitemOpen
  \bibfield  {author} {\bibinfo {author} {\bibfnamefont {E.}~\bibnamefont
  {Shchukin}}, \bibinfo {author} {\bibfnamefont {T.}~\bibnamefont {Richter}}, \
  and\ \bibinfo {author} {\bibfnamefont {W.}~\bibnamefont {Vogel}},\ }\href
  {\doibase 10.1103/PhysRevA.71.011802} {\bibfield  {journal} {\bibinfo
  {journal} {Phys. Rev. A}\ }\textbf {\bibinfo {volume} {71}},\ \bibinfo
  {pages} {011802} (\bibinfo {year} {2005})}\BibitemShut {NoStop}%
\bibitem [{\citenamefont {Mi\ifmmode~\check{s}\else \v{s}\fi{}ta}\ and\
  \citenamefont {Korolkova}(2008)}]{nk08}%
  \BibitemOpen
  \bibfield  {author} {\bibinfo {author} {\bibfnamefont {L.}~\bibnamefont
  {Mi\ifmmode~\check{s}\else \v{s}\fi{}ta}}\ and\ \bibinfo {author}
  {\bibfnamefont {N.}~\bibnamefont {Korolkova}},\ }\href {\doibase
  10.1103/PhysRevA.77.050302} {\bibfield  {journal} {\bibinfo  {journal} {Phys.
  Rev. A}\ }\textbf {\bibinfo {volume} {77}},\ \bibinfo {pages} {050302}
  (\bibinfo {year} {2008})}\BibitemShut {NoStop}%
\bibitem [{\citenamefont {Kiesel}\ and\ \citenamefont {Vogel}(2010)}]{Vogel10}%
  \BibitemOpen
  \bibfield  {author} {\bibinfo {author} {\bibfnamefont {T.}~\bibnamefont
  {Kiesel}}\ and\ \bibinfo {author} {\bibfnamefont {W.}~\bibnamefont {Vogel}},\
  }\href {\doibase 10.1103/PhysRevA.82.032107} {\bibfield  {journal} {\bibinfo
  {journal} {Phys. Rev. A}\ }\textbf {\bibinfo {volume} {82}},\ \bibinfo
  {pages} {032107} (\bibinfo {year} {2010})}\BibitemShut {NoStop}%
\bibitem [{\citenamefont {Olivares}\ and\ \citenamefont {Paris}(2011)}]{Oli11}%
  \BibitemOpen
  \bibfield  {author} {\bibinfo {author} {\bibfnamefont {S.}~\bibnamefont
  {Olivares}}\ and\ \bibinfo {author} {\bibfnamefont {M.~G.~A.}\ \bibnamefont
  {Paris}},\ }\href {\doibase 10.1103/PhysRevLett.107.170505} {\bibfield
  {journal} {\bibinfo  {journal} {Phys. Rev. Lett.}\ }\textbf {\bibinfo
  {volume} {107}},\ \bibinfo {pages} {170505} (\bibinfo {year}
  {2011})}\BibitemShut {NoStop}%
\bibitem [{\citenamefont {Tatham}\ \emph {et~al.}(2012)\citenamefont {Tatham},
  \citenamefont {Mi\ifmmode~\check{s}\else \v{s}\fi{}ta}, \citenamefont
  {Adesso},\ and\ \citenamefont {Korolkova}}]{nk12}%
  \BibitemOpen
  \bibfield  {author} {\bibinfo {author} {\bibfnamefont {R.}~\bibnamefont
  {Tatham}}, \bibinfo {author} {\bibfnamefont {L.}~\bibnamefont
  {Mi\ifmmode~\check{s}\else \v{s}\fi{}ta}}, \bibinfo {author} {\bibfnamefont
  {G.}~\bibnamefont {Adesso}}, \ and\ \bibinfo {author} {\bibfnamefont
  {N.}~\bibnamefont {Korolkova}},\ }\href {\doibase 10.1103/PhysRevA.85.022326}
  {\bibfield  {journal} {\bibinfo  {journal} {Phys. Rev. A}\ }\textbf {\bibinfo
  {volume} {85}},\ \bibinfo {pages} {022326} (\bibinfo {year}
  {2012})}\BibitemShut {NoStop}%
\bibitem [{\citenamefont {Ferraro}\ and\ \citenamefont {Paris}(2012)}]{Fer12}%
  \BibitemOpen
  \bibfield  {author} {\bibinfo {author} {\bibfnamefont {A.}~\bibnamefont
  {Ferraro}}\ and\ \bibinfo {author} {\bibfnamefont {M.~G.~A.}\ \bibnamefont
  {Paris}},\ }\href {\doibase 10.1103/PhysRevLett.108.260403} {\bibfield
  {journal} {\bibinfo  {journal} {Phys. Rev. Lett.}\ }\textbf {\bibinfo
  {volume} {108}},\ \bibinfo {pages} {260403} (\bibinfo {year}
  {2012})}\BibitemShut {NoStop}%
\bibitem [{\citenamefont {Bauke}\ and\ \citenamefont
  {Itzhak}(2011)}]{Bauke2011}%
  \BibitemOpen
  \bibfield  {author} {\bibinfo {author} {\bibfnamefont {H.}~\bibnamefont
  {Bauke}}\ and\ \bibinfo {author} {\bibfnamefont {N.~R.}\ \bibnamefont
  {Itzhak}},\ }\href@noop {} {\  (\bibinfo {year} {2011})},\ \Eprint
  {http://arxiv.org/abs/1101.2683} {arXiv:1101.2683} \BibitemShut {NoStop}%
\bibitem [{\citenamefont {Kakofengitis}\ and\ \citenamefont
  {Steuernagel}(2014)}]{ole14}%
  \BibitemOpen
  \bibfield  {author} {\bibinfo {author} {\bibfnamefont {D.}~\bibnamefont
  {Kakofengitis}}\ and\ \bibinfo {author} {\bibfnamefont {O.}~\bibnamefont
  {Steuernagel}},\ }\href@noop {} {\  (\bibinfo {year} {2014})},\ \Eprint
  {http://arxiv.org/abs/1410.4367} {arXiv:1410.4367} \BibitemShut {NoStop}%
\bibitem [{\citenamefont {Skodje}\ \emph {et~al.}(1989)\citenamefont {Skodje},
  \citenamefont {Rohrs},\ and\ \citenamefont {VanBuskirk}}]{Skodje1989a}%
  \BibitemOpen
  \bibfield  {author} {\bibinfo {author} {\bibfnamefont {R.~T.}\ \bibnamefont
  {Skodje}}, \bibinfo {author} {\bibfnamefont {H.~W.}\ \bibnamefont {Rohrs}}, \
  and\ \bibinfo {author} {\bibfnamefont {J.}~\bibnamefont {VanBuskirk}},\
  }\href {\doibase 10.1103/PhysRevA.40.2894} {\bibfield  {journal} {\bibinfo
  {journal} {Phys. Rev. A}\ }\textbf {\bibinfo {volume} {40}},\ \bibinfo
  {pages} {2894} (\bibinfo {year} {1989})}\BibitemShut {NoStop}%
\bibitem [{\citenamefont {Werner}(1988)}]{Werner1988}%
  \BibitemOpen
  \bibfield  {author} {\bibinfo {author} {\bibfnamefont {R.~F.}\ \bibnamefont
  {Werner}},\ }\href {\doibase 10.1088/0305-4470/21/24/012} {\bibfield
  {journal} {\bibinfo  {journal} {J. Phys. A}\ }\textbf {\bibinfo {volume}
  {21}},\ \bibinfo {pages} {4565} (\bibinfo {year} {1988})}\BibitemShut
  {NoStop}%
\bibitem [{\citenamefont {Dunn}\ \emph {et~al.}(1995)\citenamefont {Dunn},
  \citenamefont {Walmsley},\ and\ \citenamefont {Mukamel}}]{ian95}%
  \BibitemOpen
  \bibfield  {author} {\bibinfo {author} {\bibfnamefont {T.~J.}\ \bibnamefont
  {Dunn}}, \bibinfo {author} {\bibfnamefont {I.~A.}\ \bibnamefont {Walmsley}},
  \ and\ \bibinfo {author} {\bibfnamefont {S.}~\bibnamefont {Mukamel}},\ }\href
  {\doibase 10.1103/PhysRevLett.74.884} {\bibfield  {journal} {\bibinfo
  {journal} {Phys. Rev. Lett.}\ }\textbf {\bibinfo {volume} {74}},\ \bibinfo
  {pages} {884} (\bibinfo {year} {1995})}\BibitemShut {NoStop}%
\bibitem [{\citenamefont {Banaszek}\ and\ \citenamefont
  {W\'odkiewicz}(1996)}]{kon96}%
  \BibitemOpen
  \bibfield  {author} {\bibinfo {author} {\bibfnamefont {K.}~\bibnamefont
  {Banaszek}}\ and\ \bibinfo {author} {\bibfnamefont {K.}~\bibnamefont
  {W\'odkiewicz}},\ }\href {\doibase 10.1103/PhysRevLett.76.4344} {\bibfield
  {journal} {\bibinfo  {journal} {Phys. Rev. Lett.}\ }\textbf {\bibinfo
  {volume} {76}},\ \bibinfo {pages} {4344} (\bibinfo {year}
  {1996})}\BibitemShut {NoStop}%
\bibitem [{\citenamefont {Nogues}\ \emph {et~al.}(2000)\citenamefont {Nogues},
  \citenamefont {Rauschenbeutel}, \citenamefont {Osnaghi}, \citenamefont
  {Bertet}, \citenamefont {Brune}, \citenamefont {Raimond}, \citenamefont
  {Haroche}, \citenamefont {Lutterbach},\ and\ \citenamefont
  {Davidovich}}]{lda00}%
  \BibitemOpen
  \bibfield  {author} {\bibinfo {author} {\bibfnamefont {G.}~\bibnamefont
  {Nogues}}, \bibinfo {author} {\bibfnamefont {A.}~\bibnamefont
  {Rauschenbeutel}}, \bibinfo {author} {\bibfnamefont {S.}~\bibnamefont
  {Osnaghi}}, \bibinfo {author} {\bibfnamefont {P.}~\bibnamefont {Bertet}},
  \bibinfo {author} {\bibfnamefont {M.}~\bibnamefont {Brune}}, \bibinfo
  {author} {\bibfnamefont {J.~M.}\ \bibnamefont {Raimond}}, \bibinfo {author}
  {\bibfnamefont {S.}~\bibnamefont {Haroche}}, \bibinfo {author} {\bibfnamefont
  {L.~G.}\ \bibnamefont {Lutterbach}}, \ and\ \bibinfo {author} {\bibfnamefont
  {L.}~\bibnamefont {Davidovich}},\ }\href {\doibase
  10.1103/PhysRevA.62.054101} {\bibfield  {journal} {\bibinfo  {journal} {Phys.
  Rev. A}\ }\textbf {\bibinfo {volume} {62}},\ \bibinfo {pages} {054101}
  (\bibinfo {year} {2000})}\BibitemShut {NoStop}%
\bibitem [{\citenamefont {Lvovsky}\ \emph {et~al.}(2001)\citenamefont
  {Lvovsky}, \citenamefont {Hansen}, \citenamefont {Aichele}, \citenamefont
  {Benson}, \citenamefont {Mlynek},\ and\ \citenamefont {Schiller}}]{lvo01}%
  \BibitemOpen
  \bibfield  {author} {\bibinfo {author} {\bibfnamefont {A.~I.}\ \bibnamefont
  {Lvovsky}}, \bibinfo {author} {\bibfnamefont {H.}~\bibnamefont {Hansen}},
  \bibinfo {author} {\bibfnamefont {T.}~\bibnamefont {Aichele}}, \bibinfo
  {author} {\bibfnamefont {O.}~\bibnamefont {Benson}}, \bibinfo {author}
  {\bibfnamefont {J.}~\bibnamefont {Mlynek}}, \ and\ \bibinfo {author}
  {\bibfnamefont {S.}~\bibnamefont {Schiller}},\ }\href {\doibase
  10.1103/PhysRevLett.87.050402} {\bibfield  {journal} {\bibinfo  {journal}
  {Phys. Rev. Lett.}\ }\textbf {\bibinfo {volume} {87}},\ \bibinfo {pages}
  {050402} (\bibinfo {year} {2001})}\BibitemShut {NoStop}%
\bibitem [{\citenamefont {Allevi}\ \emph {et~al.}(2009)\citenamefont {Allevi},
  \citenamefont {Andreoni}, \citenamefont {Bondani}, \citenamefont {Brida},
  \citenamefont {Genovese}, \citenamefont {Gramegna}, \citenamefont {Traina},
  \citenamefont {Olivares}, \citenamefont {Paris},\ and\ \citenamefont
  {Zambra}}]{all09}%
  \BibitemOpen
  \bibfield  {author} {\bibinfo {author} {\bibfnamefont {A.}~\bibnamefont
  {Allevi}}, \bibinfo {author} {\bibfnamefont {A.}~\bibnamefont {Andreoni}},
  \bibinfo {author} {\bibfnamefont {M.}~\bibnamefont {Bondani}}, \bibinfo
  {author} {\bibfnamefont {G.}~\bibnamefont {Brida}}, \bibinfo {author}
  {\bibfnamefont {M.}~\bibnamefont {Genovese}}, \bibinfo {author}
  {\bibfnamefont {M.}~\bibnamefont {Gramegna}}, \bibinfo {author}
  {\bibfnamefont {P.}~\bibnamefont {Traina}}, \bibinfo {author} {\bibfnamefont
  {S.}~\bibnamefont {Olivares}}, \bibinfo {author} {\bibfnamefont {M.~G.~A.}\
  \bibnamefont {Paris}}, \ and\ \bibinfo {author} {\bibfnamefont
  {G.}~\bibnamefont {Zambra}},\ }\href {\doibase 10.1103/PhysRevA.80.022114}
  {\bibfield  {journal} {\bibinfo  {journal} {Phys. Rev. A}\ }\textbf {\bibinfo
  {volume} {80}},\ \bibinfo {pages} {022114} (\bibinfo {year}
  {2009})}\BibitemShut {NoStop}%
\bibitem [{\citenamefont {Hudson}(1974)}]{h74}%
  \BibitemOpen
  \bibfield  {author} {\bibinfo {author} {\bibfnamefont {R.~L.}\ \bibnamefont
  {Hudson}},\ }\href {\doibase 10.1016/0034-4877(74)90007-X} {\bibfield
  {journal} {\bibinfo  {journal} {Rep. Math. Phys}\ }\textbf {\bibinfo {volume}
  {6}},\ \bibinfo {pages} {294} (\bibinfo {year} {1974})}\BibitemShut {NoStop}%
\bibitem [{\citenamefont {Mari}\ and\ \citenamefont {Eisert}(2012)}]{Mari2012}%
  \BibitemOpen
  \bibfield  {author} {\bibinfo {author} {\bibfnamefont {A.}~\bibnamefont
  {Mari}}\ and\ \bibinfo {author} {\bibfnamefont {J.}~\bibnamefont {Eisert}},\
  }\href {\doibase 10.1103/PhysRevLett.109.230503} {\bibfield  {journal}
  {\bibinfo  {journal} {Phys. Rev. Lett.}\ }\textbf {\bibinfo {volume} {109}},\
  \bibinfo {pages} {230503} (\bibinfo {year} {2012})}\BibitemShut {NoStop}%
\bibitem [{\citenamefont {Veitch}\ \emph {et~al.}(2013)\citenamefont {Veitch},
  \citenamefont {Wiebe}, \citenamefont {Ferrie},\ and\ \citenamefont
  {Emerson}}]{Veitch2013}%
  \BibitemOpen
  \bibfield  {author} {\bibinfo {author} {\bibfnamefont {V.}~\bibnamefont
  {Veitch}}, \bibinfo {author} {\bibfnamefont {N.}~\bibnamefont {Wiebe}},
  \bibinfo {author} {\bibfnamefont {C.}~\bibnamefont {Ferrie}}, \ and\ \bibinfo
  {author} {\bibfnamefont {J.}~\bibnamefont {Emerson}},\ }\href {\doibase
  10.1088/1367-2630/15/1/013037} {\bibfield  {journal} {\bibinfo  {journal}
  {New J. Phys.}\ }\textbf {\bibinfo {volume} {15}},\ \bibinfo {pages} {013037}
  (\bibinfo {year} {2013})}\BibitemShut {NoStop}%
\bibitem [{\citenamefont {Pashayan}\ \emph {et~al.}(2015)\citenamefont
  {Pashayan}, \citenamefont {Wallman},\ and\ \citenamefont
  {Bartlett}}]{Pashayan2015}%
  \BibitemOpen
  \bibfield  {author} {\bibinfo {author} {\bibfnamefont {H.}~\bibnamefont
  {Pashayan}}, \bibinfo {author} {\bibfnamefont {J.~J.}\ \bibnamefont
  {Wallman}}, \ and\ \bibinfo {author} {\bibfnamefont {S.~D.}\ \bibnamefont
  {Bartlett}},\ }\href {\doibase 10.1103/PhysRevLett.115.070501} {\bibfield
  {journal} {\bibinfo  {journal} {Phys. Rev. Lett.}\ }\textbf {\bibinfo
  {volume} {115}},\ \bibinfo {pages} {070501} (\bibinfo {year}
  {2015})}\BibitemShut {NoStop}%
\bibitem [{\citenamefont {Kenfack}\ and\ \citenamefont
  {{\.{Z}}yczkowski}(2004)}]{Kenfack2004}%
  \BibitemOpen
  \bibfield  {author} {\bibinfo {author} {\bibfnamefont {A.}~\bibnamefont
  {Kenfack}}\ and\ \bibinfo {author} {\bibfnamefont {K.}~\bibnamefont
  {{\.{Z}}yczkowski}},\ }\href {\doibase 10.1088/1464-4266/6/10/003} {\bibfield
   {journal} {\bibinfo  {journal} {J. Opt. B}\ }\textbf {\bibinfo {volume}
  {6}},\ \bibinfo {pages} {396} (\bibinfo {year} {2004})}\BibitemShut {NoStop}%
\bibitem [{\citenamefont {Yearsley}\ and\ \citenamefont
  {Halliwell}(2013)}]{Yearsley2013}%
  \BibitemOpen
  \bibfield  {author} {\bibinfo {author} {\bibfnamefont {J.~M.}\ \bibnamefont
  {Yearsley}}\ and\ \bibinfo {author} {\bibfnamefont {J.~J.}\ \bibnamefont
  {Halliwell}},\ }\href {\doibase 10.1088/1742-6596/442/1/012055} {\bibfield
  {journal} {\bibinfo  {journal} {J. Phys. Conf. Ser.}\ }\textbf {\bibinfo
  {volume} {442}},\ \bibinfo {pages} {012055} (\bibinfo {year}
  {2013})}\BibitemShut {NoStop}%
\bibitem [{\citenamefont {Nicacio}\ \emph {et~al.}(2010)\citenamefont
  {Nicacio}, \citenamefont {Maia}, \citenamefont {Toscano},\ and\ \citenamefont
  {Vallejos}}]{Nicacio2010}%
  \BibitemOpen
  \bibfield  {author} {\bibinfo {author} {\bibfnamefont {F.}~\bibnamefont
  {Nicacio}}, \bibinfo {author} {\bibfnamefont {R.~N.~P.}\ \bibnamefont
  {Maia}}, \bibinfo {author} {\bibfnamefont {F.}~\bibnamefont {Toscano}}, \
  and\ \bibinfo {author} {\bibfnamefont {R.~O.}\ \bibnamefont {Vallejos}},\
  }\href {\doibase 10.1016/j.physleta.2010.08.076} {\bibfield  {journal}
  {\bibinfo  {journal} {Phys. Lett. A}\ }\textbf {\bibinfo {volume} {374}},\
  \bibinfo {pages} {4385} (\bibinfo {year} {2010})}\BibitemShut {NoStop}%
\bibitem [{\citenamefont {Yurke}\ and\ \citenamefont
  {Stoler}(1986)}]{Yurke1986}%
  \BibitemOpen
  \bibfield  {author} {\bibinfo {author} {\bibfnamefont {B.}~\bibnamefont
  {Yurke}}\ and\ \bibinfo {author} {\bibfnamefont {D.}~\bibnamefont {Stoler}},\
  }\href {\doibase 10.1103/PhysRevLett.57.13} {\bibfield  {journal} {\bibinfo
  {journal} {Phys. Rev. Lett.}\ }\textbf {\bibinfo {volume} {57}},\ \bibinfo
  {pages} {13} (\bibinfo {year} {1986})}\BibitemShut {NoStop}%
\bibitem [{\citenamefont {Dodonov}\ and\ \citenamefont
  {Man'ko}(2003)}]{dodonov2003}%
  \BibitemOpen
  \bibinfo {editor} {\bibfnamefont {V.~V.}\ \bibnamefont {Dodonov}}\ and\
  \bibinfo {editor} {\bibfnamefont {V.~I.}\ \bibnamefont {Man'ko}},\ eds.,\
  \href@noop {} {\emph {\bibinfo {title} {Theory of nonclassical states of
  light}}}\ (\bibinfo  {publisher} {Taylor \& Francis},\ \bibinfo {address}
  {London New York},\ \bibinfo {year} {2003})\BibitemShut {NoStop}%
\bibitem [{\citenamefont {Barnett}\ and\ \citenamefont
  {Radmore}(1997)}]{Barnett1997}%
  \BibitemOpen
  \bibfield  {author} {\bibinfo {author} {\bibfnamefont {S.~M.}\ \bibnamefont
  {Barnett}}\ and\ \bibinfo {author} {\bibfnamefont {P.~M.}\ \bibnamefont
  {Radmore}},\ }\href@noop {} {\emph {\bibinfo {title} {{Methods in theoretical
  quantum optics}}}}\ (\bibinfo  {publisher} {Oxford University Press},\
  \bibinfo {address} {Oxford, New York},\ \bibinfo {year} {1997})\BibitemShut
  {NoStop}%
\bibitem [{\citenamefont {L{\"{u}}tkenhaus}\ and\ \citenamefont
  {Barnett}(1995)}]{Lutkenhaus1995}%
  \BibitemOpen
  \bibfield  {author} {\bibinfo {author} {\bibfnamefont {N.}~\bibnamefont
  {L{\"{u}}tkenhaus}}\ and\ \bibinfo {author} {\bibfnamefont {S.~M.}\
  \bibnamefont {Barnett}},\ }\href {\doibase 10.1103/PhysRevA.51.3340}
  {\bibfield  {journal} {\bibinfo  {journal} {Phys. Rev. A}\ }\textbf {\bibinfo
  {volume} {51}},\ \bibinfo {pages} {3340} (\bibinfo {year}
  {1995})}\BibitemShut {NoStop}%
\bibitem [{Note1()}]{Note1}%
  \BibitemOpen
  \bibinfo {note} {Other generalized distributions in phase space that satisfy
  this property exist~\cite {Cohen1966a,Loughlin2003}, but they are not
  relevant for the topic discussed here.}\BibitemShut {Stop}%
\bibitem [{\citenamefont {Vacchini}(2002)}]{Vacchini2002}%
  \BibitemOpen
  \bibfield  {author} {\bibinfo {author} {\bibfnamefont {B.}~\bibnamefont
  {Vacchini}},\ }\href {\doibase 10.1063/1.1505126} {\bibfield  {journal}
  {\bibinfo  {journal} {J. Math. Phys.}\ }\textbf {\bibinfo {volume} {43}},\
  \bibinfo {pages} {5446} (\bibinfo {year} {2002})}\BibitemShut {NoStop}%
\bibitem [{\citenamefont {Yearsley}(2010)}]{Yearsley2010}%
  \BibitemOpen
  \bibfield  {author} {\bibinfo {author} {\bibfnamefont {J.~M.}\ \bibnamefont
  {Yearsley}},\ }\href {\doibase 10.1103/PhysRevA.82.012116} {\bibfield
  {journal} {\bibinfo  {journal} {Phys. Rev. A}\ }\textbf {\bibinfo {volume}
  {82}},\ \bibinfo {pages} {012116} (\bibinfo {year} {2010})}\BibitemShut
  {NoStop}%
\bibitem [{\citenamefont {Paavola}\ \emph {et~al.}(2011)\citenamefont
  {Paavola}, \citenamefont {Hall}, \citenamefont {Paris},\ and\ \citenamefont
  {Maniscalco}}]{Paavola2011a}%
  \BibitemOpen
  \bibfield  {author} {\bibinfo {author} {\bibfnamefont {J.}~\bibnamefont
  {Paavola}}, \bibinfo {author} {\bibfnamefont {M.~J.~W.}\ \bibnamefont
  {Hall}}, \bibinfo {author} {\bibfnamefont {M.~G.~A.}\ \bibnamefont {Paris}},
  \ and\ \bibinfo {author} {\bibfnamefont {S.}~\bibnamefont {Maniscalco}},\
  }\href {\doibase 10.1103/PhysRevA.84.012121} {\bibfield  {journal} {\bibinfo
  {journal} {Phys. Rev. A}\ }\textbf {\bibinfo {volume} {84}},\ \bibinfo
  {pages} {012121} (\bibinfo {year} {2011})}\BibitemShut {NoStop}%
\bibitem [{\citenamefont {Cohen}(1966)}]{Cohen1966a}%
  \BibitemOpen
  \bibfield  {author} {\bibinfo {author} {\bibfnamefont {L.}~\bibnamefont
  {Cohen}},\ }\href {\doibase 10.1063/1.1931206} {\bibfield  {journal}
  {\bibinfo  {journal} {J. Math. Phys.}\ }\textbf {\bibinfo {volume} {7}},\
  \bibinfo {pages} {781} (\bibinfo {year} {1966})}\BibitemShut {NoStop}%
\bibitem [{\citenamefont {Loughlin}\ and\ \citenamefont
  {Cohen}(2003)}]{Loughlin2003}%
  \BibitemOpen
  \bibfield  {author} {\bibinfo {author} {\bibfnamefont {P.~J.}\ \bibnamefont
  {Loughlin}}\ and\ \bibinfo {author} {\bibfnamefont {L.}~\bibnamefont
  {Cohen}},\ }\href {\doibase 10.1080/09500340308233563} {\bibfield  {journal}
  {\bibinfo  {journal} {J. Mod. Opt.}\ }\textbf {\bibinfo {volume} {50}},\
  \bibinfo {pages} {2305} (\bibinfo {year} {2003})}\BibitemShut {NoStop}%
\end{thebibliography}%
\end{document}